\documentclass[twocolumn, floatfix]{aastex631}
\usepackage[utf8]{inputenc} 
\usepackage[T1]{fontenc}   
\usepackage{hyperref}    
\usepackage{silence}
\usepackage{url}       
\usepackage{float}
\usepackage{booktabs}  
\usepackage{fontawesome}
\usepackage{amsfonts}       
\usepackage{amsmath, amssymb}
\usepackage{bm}
\usepackage{nicefrac}  
\usepackage{microtype}
\usepackage{xcolor}
\usepackage{wrapfig} 
\usepackage{aasmacros}  
\definecolor{Red}               {RGB}{255, 0, 0}        
\definecolor{Blue}              {RGB}{0, 24, 168}
\definecolor{Purple}            {cmyk}{0.4, 1, 0., 0.05}
\definecolor{White}            {RGB}{255, 255, 255}
\definecolor{Lavender}         {RGB}{57, 21, 105}
\usepackage{tikz}
\usetikzlibrary{calc}
\usepackage{physics}
\hypersetup{
    colorlinks,
    linkcolor={red!50!black},
    citecolor={blue!50!black},
    urlcolor={blue!80!blue}
}

\DeclareRobustCommand{\bbone}{\text{\usefont{U}{bbold}{m}{n}1}}

\newcommand{\vis}{\mathcal{V}}

\begin{document}
\title{
    \centering
    IRIS: A Bayesian Approach for Image Reconstruction in Radio Interferometry \\
    with expressive Score-Based priors.  
}

\author[0009-0009-6353-0950]{Noé Dia}
\affiliation{Université de Montréal, Montréal, Canada}
\affiliation{Mila - Quebec Artificial Intelligence Institute, Montréal, Canada}
\affiliation{Ciela - Montreal Institute for Astrophysical Data Analysis and Machine Learning, Montréal, Canada}

\author[0000-0001-5200-4095]{M. J. Yantovski-Barth}
\affiliation{Université de Montréal, Montréal, Canada}
\affiliation{Mila - Quebec Artificial Intelligence Institute, Montréal, Canada}
\affiliation{Ciela - Montreal Institute for Astrophysical Data Analysis and Machine Learning, Montréal, Canada}

\author[0000-0001-8806-7936]{Alexandre Adam}
\affiliation{Université de Montréal, Montréal, Canada}
\affiliation{Mila - Quebec Artificial Intelligence Institute, Montréal, Canada}
\affiliation{Ciela - Montreal Institute for Astrophysical Data Analysis and Machine Learning, Montréal, Canada}

\author[0000-0001-5838-8405]{Micah Bowles}
\affiliation{Department of Astrophysics, University of Oxford, Oxford UK}

\author[0000-0003-3544-3939]{Laurence Perreault-Levasseur}
\affiliation{Université de Montréal, Montréal, Canada}
\affiliation{Mila - Quebec Artificial Intelligence Institute, Montréal, Canada}
\affiliation{Ciela - Montreal Institute for Astrophysical Data Analysis and Machine Learning, Montréal, Canada}
\affiliation{Flatiron Institute - Center for Computational Astrophysics, New York, USA}
\affiliation{Perimeter Institute, Waterloo, Canada}
\affiliation{Trottier Space Institute, McGill University, Montréal, Canada}

\author[0000-0002-8669-5733]{Yashar Hezaveh}
\affiliation{Université de Montréal, Montréal, Canada}
\affiliation{Mila - Quebec Artificial Intelligence Institute, Montréal, Canada}
\affiliation{Ciela - Montreal Institute for Astrophysical Data Analysis and Machine Learning, Montréal, Canada}
\affiliation{Flatiron Institute - Center for Computational Astrophysics, New York, USA}
\affiliation{Perimeter Institute, Waterloo, Canada}
\affiliation{Trottier Space Institute, McGill University, Montréal, Canada}

\author[0000-0002-5364-2301]{Anna Scaife}
\affiliation{University of Manchester, Manchester, UK}
\affiliation{The Alan Turing Institute, UK}

\received{XXXX}
\revised{YYYY}
\accepted{ZZZZ}
\submitjournal{The Astrophysical Journal}
\begin{abstract}
Inferring sky surface brightness distributions from noisy interferometric data in a principled statistical framework has been a key challenge in radio astronomy. In this work, we introduce \textbf{I}maging for \textbf{R}adio \textbf{I}nterferometry with \textbf{S}core-based models (IRIS). We use score-based models trained on optical images of galaxies as an expressive prior in combination with a Gaussian likelihood in the $uv$-space to infer images of protoplanetary disks from visibility data of the DSHARP survey conducted by ALMA. We demonstrate the advantages of this framework compared with traditional radio interferometry imaging algorithms, showing that it produces plausible posterior samples despite the use of a misspecified galaxy prior. Through coverage testing on simulations, we empirically evaluate the accuracy of this approach to generate calibrated posterior samples. Our code is open source and freely available at \texttt{https://github.com/EnceladeCandy/IRIS}.
\end{abstract}

\section{Introduction}
Interferometry is a technique that combines signals received by an array of antennae to probe scales at a resolution higher than what is attainable by individual single-dish telescopes. From the first astronomical observations with a two-element interferometer \citep{Ryle1952} to the first-ever image of a black hole by the Event Horizon Telescope Collaboration \citep{EventHorizonCollab2019}, radio interferometry has a rich record of successes that have enabled advances in our understanding of the Universe.   

Unlike CCDs that directly measure surface brightness at the focal plane of telescopes, interferometry measures the Fourier modes of the sky at Fourier coordinates ($uv$-coordinates) determined by the separation between the antennae of the array.
One of the major challenges in producing an image from interferometric data is the incomplete sampling of Fourier space.
An inverse Fourier transform of the data will result in what is referred to as the \textit{dirty image}, where substantial artifacts are present depending on the specific antenna configuration. 
As the sky surface brightness is not directly measured by interferometry, it should be inferred from noisy Fourier modes (visibilities), a process known as \textit{synthesis imaging}. Although the Earth rotation \citep{Ryle1962} and multi-frequency synthesis \citep{Sault1999} alleviate the problem of sparsity of the $uv$-coverage, recovering a sky brightness from measured visibilities remains an ill-posed inverse problem. 
 
The most widely-used imaging methods are derivatives of CLEAN \citep{Hogbom1974, Cornwell2008multiscale, Rau2011, Offringa2017,clean_review_2009}, a deconvolution algorithm which attempts to iteratively remove the effects of the incomplete uv-space sampling from the dirty image. The original CLEAN approach attempts to address this by operating under an assumption of how the structure of the emission can be decomposed into Gaussian or point sources. This can introduce significant limitations when imaging extended or complex objects; the algorithm can produce images with unphysical negative flux and results in suboptimal resolution. CLEAN also requires expert knowledge to determine the optimal parameters yielding the best reconstruction. As an alternative, the MEM algorithm \citep{Cornwell2008multiscale} incorporates prior knowledge about the sky brightness in its maximum entropy approach, aiming to produce smoother images without the need for extensive manual intervention. MEM however struggles with the reconstruction of point sources, especially when superimposed on a smooth background \citep[see Sec 11.2.2 of][]{Thompson2017}. More importantly, neither CLEAN nor MEM provide accurate uncertainty estimates \citep{2021_bayesian_vs_clean}.

These limitations have motivated efforts to develop alternative imaging algorithms. MPoL \citep{mpol}, for example, adopts a Regularized Maximum Likelihood approach. Still, this method does not provide uncertainty quantification and relies on the fine-tuning of regularization parameters (a prior in a Bayesian interpretation). Another approach is \texttt{resolve} \citep{Junwlewitz2016, Arras2018, Arras2019, knollmüller2020}, which adopts a log-normal prior to make the problem of Bayesian inference tractable. Numerous other approaches have been developed \citep{Morningstar2018, Schmidt2022, Wang2023,  Drozdova2024, R2D2, AIRI, PolyCLEAN, QuantifAI} to tackle the task of radio interferometric imaging but few are applied to real data or offer the statistical advantages inherent to Bayesian methods.

In recent years, score models have emerged as a powerful method to encode probability distributions in high-dimensional spaces. In combination with a likelihood term, these models have been used to perform Bayesian inference over images \citep{Remy2022,Song2022, Chung2022, Karchev2022, Adam2022, Adam2023, Xue2023, Barco2024}, including in the context of interferometric image synthesis \citep{Wang2023,Feng2023a, Feng2023b, feng2024eventhorizonscaleimagingm87different,  Drozdova2024}, using a sampling procedure referred to as diffusion. 
The main advantages of this framework are that it can encode flexible priors and efficiently produce samples from posteriors in high dimensional spaces.

In this work, we use score-based generative models \citep{Ho2020,Song_sbm_2021} as data-driven priors \citep{Graikos2022} to perform Bayesian inference for radio interferometric imaging.  
We test the inference pipeline on simulations drawn from the learned priors and through coverage testing show that it can produce unbiased samples.
We then apply the approach to ALMA observations of protoplanetary disks from the Disk Substructures at High Angular Resolution Project (DSHARP) survey \citep{dsharp1_2018}. 

The outline of the paper is as follows. In Section \ref{sec:methods}, we describe the general approach to Bayesian inference for radio interferometric imaging by introducing the framework of score-based modeling through Stochastic Differential Equations (SDEs). In Section \ref{sec:data_and_phys}, we detail the physical model and the data used in this work, including the training data for our machine learning model as well as the ALMA observations of protoplanetary disks. In Section \ref{sec:sims}, we test the performance of the proposed approach on simulations and finally apply it to ALMA data in Section \ref{sec:results}. 

\section{Score-based models for posterior inference}\label{sec:methods}
Radio interferometers measure visibilities. For distant sources, the van Cittert-Zernike theorem \citep{vancittert1934, Zernike1938} states that these visibilities correspond to the Fourier modes of the true sky emission. Due to the finite amount of antennae available, only a subsampled set of visibilities is measured. Moreover, visibilities are subject to additive instrumental noise. For ALMA, the noise in the real and imaginary components of the visibilities' errors are modeled as independent realizations of a zero-centered diagonal Gaussian. Due to the combination of these two factors --- incomplete and noisy measurements--- recovering the sky brightness given a set of observed visibilities constitutes a noisy and ill-posed inverse problem.

As such, our goal is to recover plausible surface brightness profiles over a pixel grid, $\mathbf{x} \in \mathbb{R}^n$, where $n$ is the number of pixels of our model, using ALMA observations of a protoplanetary disk, $\tilde{\vis} \in \mathbb{C}^{m}$, where $m$ is the number of gridded visibilities sampled (we expand on the gridding process in section \ref{sec:data_and_phys}). The measurement equation of a radio interferometer can be written as 
\begin{equation}\label{eq:vis}
    \tilde{\vis} = \tilde{A} \mathbf{x} + \tilde{\boldsymbol{\eta}} \, ,
\end{equation}
where $\tilde{A} \in \mathbb{C}^{m \times n}$ is a matrix representing the linear physical model mapping the sky brightness $\mathbf{x}$ to the complex visibilities $\tilde{\mathcal{V}}$ and $\tilde{\boldsymbol{\eta}}\sim \mathcal{CN}(0, \tilde{\Sigma})$ is additive instrumental noise where $\mathcal{CN}(\cdot)$ denotes a complex Gaussian with a complex diagonal covariance matrix $\tilde{\Sigma}\in \mathbb{C}^{m\times m}$. In order to treat the complex Gaussian likelihood involved by the data generating process Eq. \eqref{eq:vis}, we use throughout this work a \textit{vectorized representation} of complex variables. In this construction, every complex random variable is reformulated by concatenating the real and the imaginary part, e.g. $\tilde{\mathbf{z}}\in \mathbb{C}^{m}$ is represented as $\mathbf{z} = \left(\text{Re}(\tilde{\mathbf{z}}), \text{Im}(\bar{\mathbf{z}}) \right)$. For complex matrices, such as $\tilde{A}$, we construct a corresponding real-valued block matrix $A \in \mathbb{R}^{2m \times n}$, 
\begin{align}
    A = 
    \begin{pmatrix}
        \text{Re}(\tilde{A}) \\
        \text{Im}(\tilde{A}) 
    \end{pmatrix} \, .
\end{align} 

In a Bayesian inference setting, solving this inverse problem translates into the task of sampling from the posterior defined by Bayes's theorem as the product of the likelihood, $p(\vis \mid \mathbf{x}) = p(\boldsymbol{\eta})$, which encodes the properties of the additive noise distribution, and the prior, $p(\mathbf{x})$:
\begin{equation}\label{eq:Bayes}
    p(\mathbf{x} \mid \vis) \propto p(\vis \mid \mathbf{x}) p(\mathbf{x})\, .
\end{equation}
The prior represents the known information about the uncertain parameters, here the sky emission, $\mathbf{x}$, before considering any measurement. 
We summarize herein our methodology to learn the prior, characterize the likelihood, and to sample from the posterior.

\subsection{Learning a prior} 

Generative models aim to encode the underlying probability distribution of data. Score-based models (SBMs) have recently emerged as the state-of-the-art class of models to perform this task, especially for high-dimensional probability distributions. An SBM uses a neural network $\mathbf{s}_{\theta}(\mathbf{x}): \mathbb{R}^{n} \rightarrow \mathbb{R}^{n}$ to explicitly learn the score of a probability distribution, $\nabla_{\mathbf{x}}\log p(\mathbf{x})$, and to approximate it such that at the end of the training procedure $\mathbf{s}_{\theta}(\mathbf{x}) \approx \grad_{\mathbf{x}} \log p(\mathbf{x})$. 
The score is then used to generate new samples from $p(\mathbf{x})$ by starting from samples from a Gaussian distribution and using the score to guide a stochastic process towards samples from the target distribution, i.e. to sample a prior distribution close to the dataset of images $\mathcal{D}$ on which the SBM was trained. 
The stochastic process used in this work follows the framework based on Stochastic Differential Equations (SDEs) as introduced by \cite{Song_sbm_2021}. 

\subsubsection{Score Matching}
Modeling the score of a prior distribution (i.e. learning it) without access to the true score is known as Score Matching \citep{Hyvarinen2005, Vincent2011, Song2019}. In this work, we use Denoising Score Matching (DSM) to learn a time-dependent score $\nabla_{\mathbf{x}_t} \log p_t(\mathbf{x}_t)$ over different noise scales $\sigma(t)$ (also called temperatures) where $t\in [0, 1]$ is the time variable of an underlying SDE. The subscript $t$ in the notation $p_t(\cdot)$ indicates that the density function's statistics are dependent on time; similarly, samples of the density function $p_t(\cdot)$ are noted $\mathbf{x}_t$. For our neural network, we choose a U-Net architecture \citep{Ronneberger2015} and condition it on the time index $t$. We also follow previous works \citep{Song2020} by predicting the quantity $\mathbf{s}_{\theta}(\mathbf{x}_t,t) = \boldsymbol{\epsilon}_{\theta} (\mathbf{x}_t,t)/ \sigma(t)$ where $\boldsymbol{\epsilon}_{\theta}(\cdot)$ is the SBM. The DSM translates into a minimization objective of the weighted sum of the Fisher divergence between the score model, $\mathbf{s}_{\theta}(\mathbf{x}_t,t)$, and the score of a Gaussian perturbation kernel, $p(\mathbf{x}_t \mid \mathbf{x}_0)=\mathcal{N}(\mathbf{x}_t \mid \mu(t)\mathbf{x}_0, \sigma^2(t)\bbone)$, where $\mu(t)$ and $\sigma(t)$ are scalar functions dependent on the SDE. The training objective can therefore be written as
\begin{align}\label{eq:training_objective}
\mathcal{L}_{\theta} = 
    \underset{\substack{
    \mathbf{x}_0\sim \mathcal{D}\\
    t\sim\mathcal{U}(0, 1)\\ 
    \mathbf{x}_t \sim p(\mathbf{x}_t \mid \mathbf{x}_0)
    }}{
    \mathbb{E}}
    \left[ 
    \lambda(t) \big\lVert \mathbf{s}_{\theta} (\mathbf{x}_t, t) - \grad_{\mathbf{x}_t} \log p (\mathbf{x}_t \mid\mathbf{x}_0) \big\rVert^{2} 
    \right] \, ,
\end{align}
where $\lambda(t)$ is a weighting term. As shown in \cite{Song_sbm_2021, Song2021ml}, a careful choice of $\lambda(t)$ according to the SDE reformulates Eq. \eqref{eq:training_objective} as a minimization objective of the Kullback-Leibler (KL) divergence between the distribution learned by the SBM and the target distribution. We use the \texttt{score\_models}\footnote{\href{https://github.com/AlexandreAdam/score_models}{\faGithub \vspace{0.2cm} github.com/AlexandreAdam/score\_models}} package to fit the score function $\mathbf{s}_{\theta}(\mathbf{x}_t,t)$.

\subsubsection{Forward SDE}
The DSM objective can be interpreted within the framework of SDEs. In this work, we consider a Ornstein-Uhlenbeck process as the forward SDE to add noise to the samples during training: 
\begin{align}\label{eq:forward_sde}
    d\mathbf{x}_t = -\frac{1}{2}\beta(t)\mathbf{x}_t dt + g(t) d \mathbf{w}_t \, , 
\end{align}
where $\beta(t): \mathbb{R} \rightarrow \mathbb{R}$ is the drift coefficient, $g(\cdot):\mathbb{R}\rightarrow \mathbb{R}$ is an homogeneous diffusion coefficient and $\mathbf{w}_t\in \mathbb{R}^{n}$ is a Wiener process. In the context of the density function, the forward process is equivalent to convolving the target distribution $p_0 (\mathbf{x}_0)$ with a perturbation kernel $p(\mathbf{x}_t \mid \mathbf{x}_0)$, a process known as \textit{annealing}. Since the kernel is Gaussian for the SDEs used in this work, an image at temperature $t$, $\mathbf{x}_t$, can be obtained simply by adding Gaussian noise to an input image $\mathbf{x}_0$: 
\begin{align}\label{eq:noising}
    \mathbf{x}_t = \mu(t)\mathbf{x}_0 + \sigma(t)\mathbf{z} \, ,
\end{align}
where $\mathbf{z}\sim \mathcal{N}(0, \bbone_{n \times n})$ and $\mu(t)=e^{-\frac{1}{2}\int_{0}^{t}\beta(s)ds}$. Note that this $\mathbf{x}_t$ is exactly the term that the SBM takes as input during training Eq.~\eqref{eq:training_objective}.

\subsubsection{Reverse SDE}
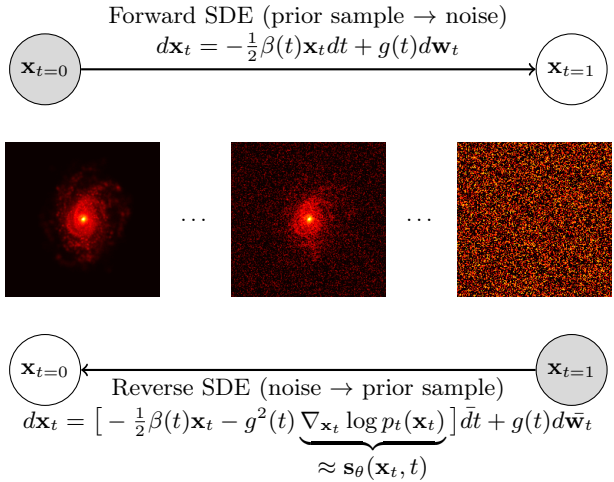
\begin{figure}
    \begin{tikzpicture}
    \node (start) at (1, 0) [circle, draw, fill=gray!30] {$\mathbf{x}_{t=0}$};
    \node (end) at (8, 0) [circle, draw, fill=white!30] {$\mathbf{x}_{t=1}$};
    \node (startRev) at (1, -4) [circle, draw, fill=white!30] {$\mathbf{x}_{t=0}$};
    \node (endRev) at (8, -4) [circle, draw, fill=gray!30] {$\mathbf{x}_{t=1}$};
    \node at (4.5, 0.7) {Forward SDE (prior sample $\rightarrow$ noise)};
    \draw[->, thick] (start) -- (end);
    \node at (4.5, 0) [above] {$d\mathbf{x}_t = -\frac{1}{2}\beta(t)\mathbf{x}_t dt + g(t) d\mathbf{w}_t$};
    \node[inner sep=0pt] (image1) at (1.5, -2.) {\includegraphics[scale=0.4]{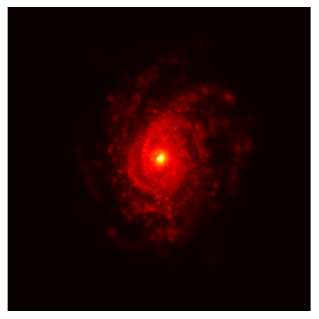}};
    \node[inner sep=0pt] (image2) at (4.5, -2.) {\includegraphics[scale=0.4]{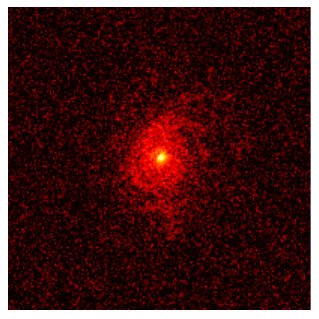}};
    \node[inner sep=0pt] (image3) at (7.5, -2.) {\includegraphics[scale=0.4]{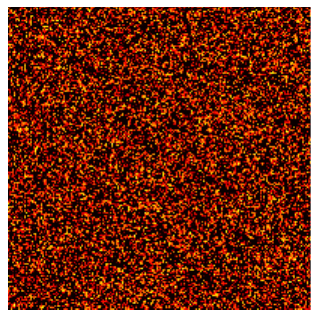}};
    \node at (4.5, -4.25) {Reverse SDE (noise $\rightarrow$ prior sample)};
    \draw[->, thick] (endRev) -- (startRev);
    \node at (4.5, -5){
        $d\mathbf{x}_t = \big[-\frac{1}{2}\beta(t)\mathbf{x}_t - g^2(t) \underbrace{\nabla_{\mathbf{x}_t} \log p_t(\mathbf{x}_t)}_{\displaystyle \approx \mathbf{s}_{\theta}(\mathbf{x}_t, t)}\big] \bar{dt} + g(t) d\bar{\mathbf{w}_t}$
    };
    \node at (3, -2) {\footnotesize $\dots$};
        \node at (6, -2) {\footnotesize$\dots$}; 
    \end{tikzpicture}
    \caption{The Forward SDE and the Reverse SDE for a score model trained on galaxy images (SKIRT dataset). The starting point of each process is colored in gray.  Prior samples are generated by starting from Gaussian noise and solving the reverse SDE while approximating the score function by a SBM.}
    \label{fig:prior_sampling}
\end{figure}

A forward SDE has a corresponding \textit{reverse SDE}, where time is reversed  while ensuring to preserve the underlying dynamics of the stochastic process. Sampling from the target distribution $p_0(\mathbf{x}_0)$ involves solving the reverse SDE using Anderson's reverse-time formula \citep{Anderson1982}
\begin{align}\label{eq:reverse_sde}
    d \mathbf{x}_t &= \left[-\frac{1}{2}\beta(t)\mathbf{x}_t - g^{2}(t) \grad_{\mathbf{x}_t}\log p_t( \mathbf{x}_t) \right]\bar{dt} + g(t) d \bar{\mathbf{w}}_t\, ,
\end{align}
where $\bar{\mathbf{w}}_t$ is a time-reversed Wiener process and $\bar{dt}$ is a negative infinitesimal time step. In this work, we use two specific SDEs, the Variance Exploding (VE) and the Variance Preserving (VP) SDE introduced by \cite{Song_sbm_2021} (refer to Appendix \ref{app:sdes} for more details on these two SDEs). Eq. \eqref{eq:reverse_sde} can be iteratively solved using discretization techniques, such as the Euler-Maruyama method or Predictor-Corrector (PC) approaches \citep{Hairer2010}. We explore both samplers for our calibration tests (see Sec. \ref{sec:sims}). 
By substituting the score function with a trained SBM $\mathbf{s}_{\theta}(\mathbf{x}_t, t)$, one can sample the target distribution with a good approximation. We show the forward and reverse process of the SDE for a SBM trained on a dataset of galaxy images Figure~\ref{fig:prior_sampling}.

The formalism presented above can be interpreted with a Bayesian view to learn an expressive prior over high-dimensional probability distributions such as real or simulated images of astrophysical objects.

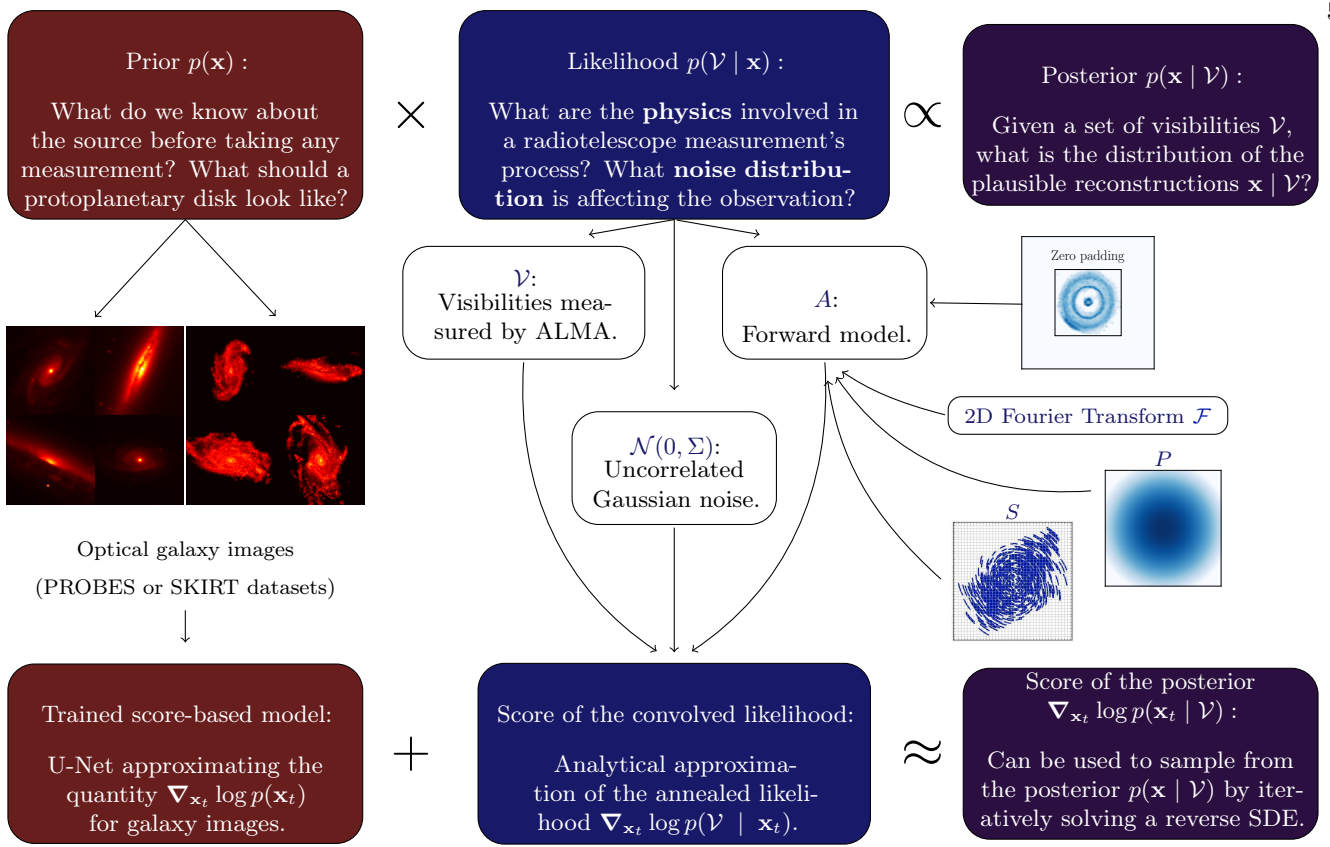
\begin{figure*}
\centering
    \vspace{-0.5cm}
    \begin{tikzpicture}
    \definecolor{tikzred}{RGB}{105, 30, 30}
    \definecolor{tikzblue}{RGB}{25, 25, 105}
    \definecolor{tikzpurple}{RGB}{43, 15, 64}
        \node[draw,rounded corners=10pt,minimum width=1cm,minimum height=2cm,text width=4.5cm,align=center,fill=tikzred] (box1) at (-6,0) {\begin{center}\textcolor{White}{Prior} $\color{White} p(\mathbf{x}):$\end{center}
         \textcolor{White}{What do we know about the source before taking any measurement? What should a protoplanetary disk look like?} };
        \node[draw,rounded corners=10pt,minimum width=1cm,minimum height=2cm,text width=5.5cm,align=center, fill=tikzblue] (box2) at (0.5,0) {\begin{center}\textcolor{White}{Likelihood} $\color{White}  p(\mathcal{V}\mid \mathbf{x}):$ \end{center}
        \textcolor{White}{What are the \textbf{physics} involved in a radiotelescope measurement's process? What \textbf{noise distribution} is affecting the observation?}};
        \node at (-3., 0) {\huge$\times$}; 
        \node at (-3, -8.5) {\huge $+$};
        \node[draw,rounded corners=10pt,minimum width=1cm,minimum height=2cm,text width=4.5cm,align=center, fill=tikzpurple] (box3) at (6.7,0) {\begin{center}\textcolor{White}{Posterior} $\color{White} p(\mathbf{x}\mid\mathcal{V}):$\end{center}
        \textcolor{White}{Given a set of visibilities $\mathcal{V}$, what is the distribution of the plausible reconstructions $\color{White} \mathbf{x} \mid \mathcal{V}$?}};
        \node at (3.8, 0) {\huge$\propto$};
        \node at (3.8, -8.5) {\huge$\approx$};
        \node[draw,rounded corners=10pt,minimum width=1cm,minimum height=2cm,text width=4.5cm,align=center, fill=tikzred] (box4) at (-6, -8.5) {
        \begin{center}\textcolor{White}{Trained score-based model:}\end{center}
        \textcolor{White}{U-Net approximating the quantity 
        $\color{White} \grad_{\mathbf{x}_{t}} \log p(\mathbf{x}_t)$ for galaxy images.}};
        \node[draw,rounded corners=10pt,minimum width=1cm,minimum height=2cm,text width=5.cm,align=center, fill=tikzblue] (box5) at (0.5, -8.5) {
        \begin{center}\textcolor{White}{Score of the convolved likelihood:}\end{center} \textcolor{White}{Analytical approximation 
        of the annealed likelihood $\color{White} \grad_{\mathbf{x}_{t}} \log p(\mathcal{V}\mid\mathbf{x}_t)$.
        }};
        \node[draw,rounded corners=10pt,minimum width=1cm,minimum height=2cm,text width=4.5cm,align=center, fill=tikzpurple] (box6) at (6.7,-8.5) {\vspace{-0.5cm} \begin{center}
        \textcolor{White}{Score of the posterior \\}$\color{White} \grad_{\mathbf{x}_{t}} \log p(\mathbf{x}_{t}\mid \mathcal{V}):$
        \end{center}  
        \textcolor{White}{Can be used to sample from the posterior 
        $\color{White} p(\mathbf{x}\mid\mathcal{V})$ by iteratively solving a reverse SDE.}};
        \node (image1) at (-7.2,-4) {\includegraphics[scale = 0.1]{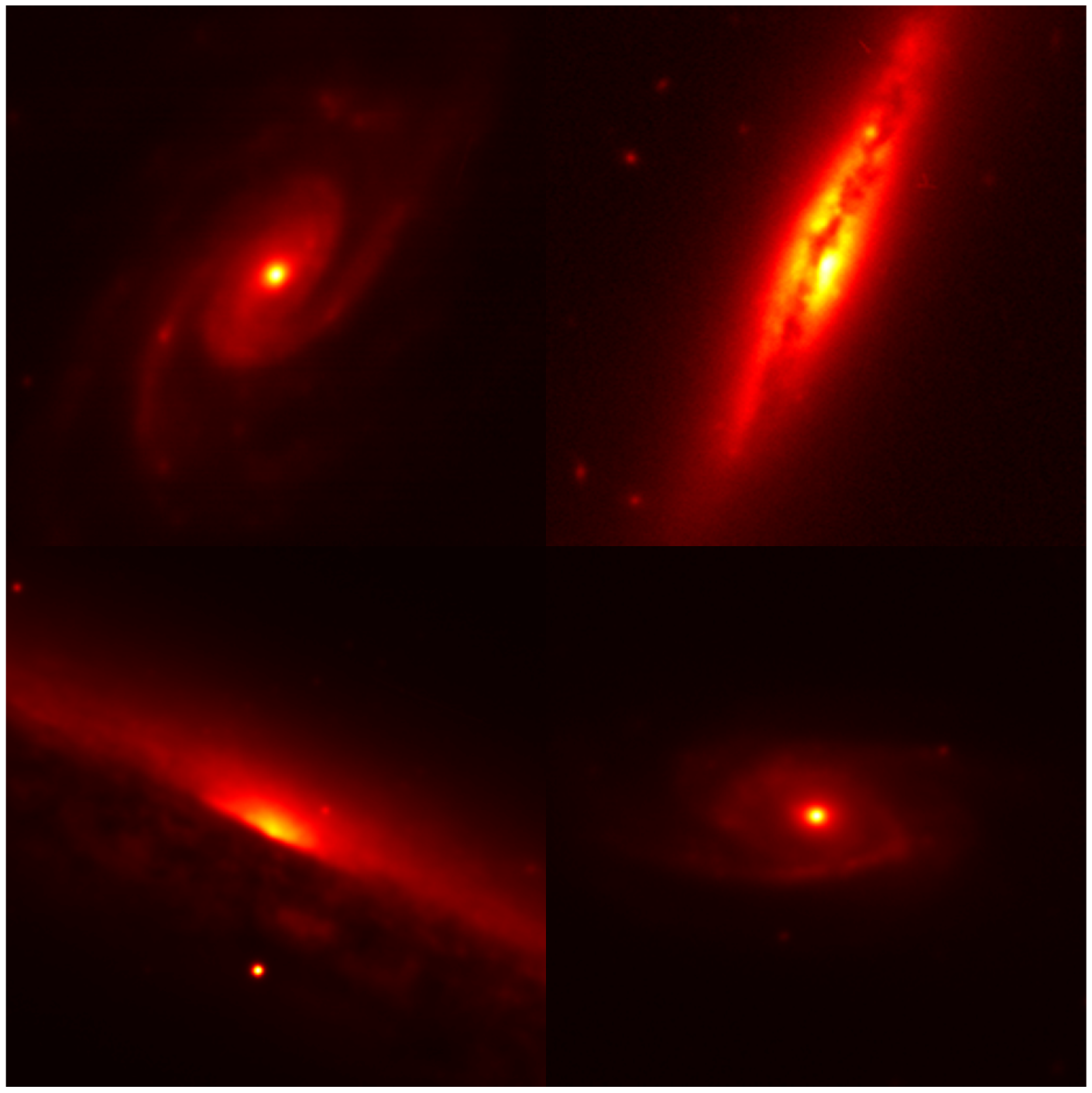}};
        \node (image2) at (-4.8,-4) {\includegraphics[scale = 0.1]{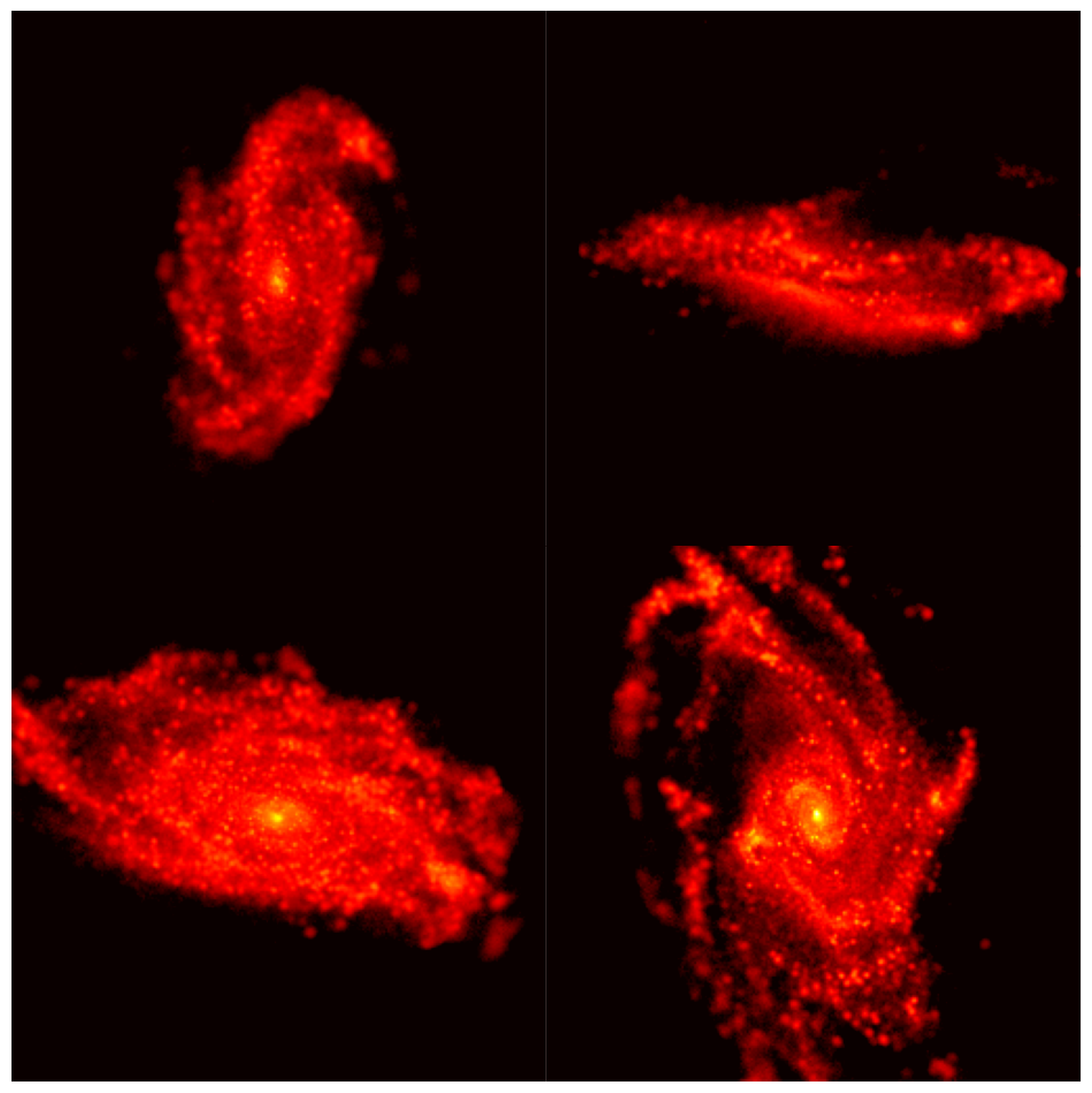}};
        \node (title1) at (-6, -5.8) {\footnotesize Optical galaxy images};
        \node (title1) at (-6, -6.3) {\footnotesize (PROBES or SKIRT datasets)};
        \node[draw,rounded corners=10pt,minimum width=1.5cm,minimum height=1.5cm,text width=3cm,align=center] (V) at (-1.5,-2.5) {\vspace{-0.35cm}\begin{center} $\color{tikzblue} \mathcal{V}$:\end{center}\vspace{-0.35cm}
        Visibilities measured by ALMA.};
        \node[draw,rounded corners=10pt,minimum width=1cm,minimum height=1.5cm,text width=2.5cm,align=center] (A) at (2.5,-2.5) {\begin{center} $\color{tikzblue} A$:\end{center}\vspace{-0.2cm} Forward model.\vspace{0.1cm}};
        \node[draw,rounded corners=10pt,minimum width=1cm,minimum height=1.5cm,text width=2.5cm,align=center] (N) at (0.5,-4.7) {\vspace{-0.35cm}\begin{center} $\color{tikzblue} \mathcal{N}(0, \Sigma)$:\end{center}\vspace{-0.35cm}
        Uncorrelated Gaussian  noise.};
        \node (image4) at (6,-2.5) {\includegraphics[scale = 0.15]{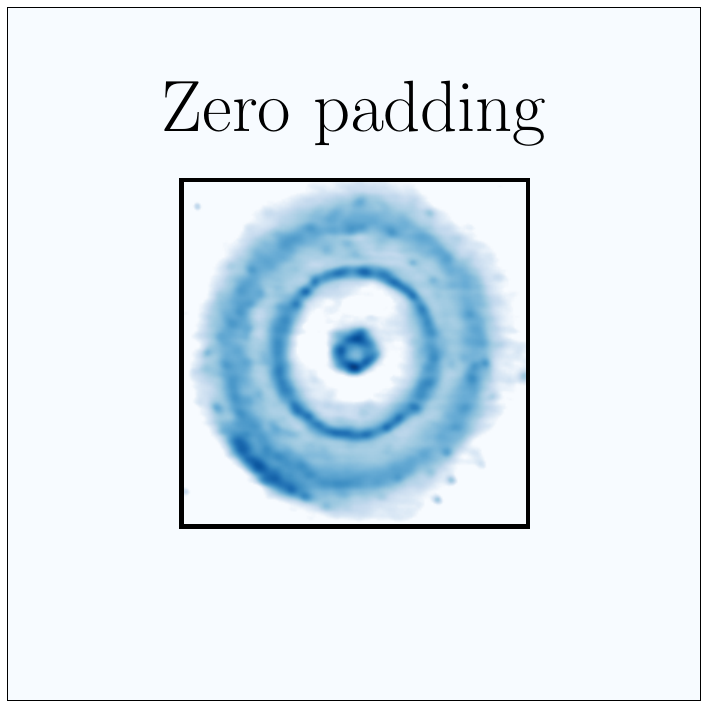}};
        \node[draw,rounded corners=5pt, minimum width=1cm,minimum height=0.5cm,text width=3.5cm,align=center] (fourier) at (6., -4.) {\footnotesize \textcolor{tikzblue}{2D Fourier Transform} $\color{Blue} \mathcal{F}$};
        \node (image6) at (7,-5.5) {\includegraphics[scale = 0.3]{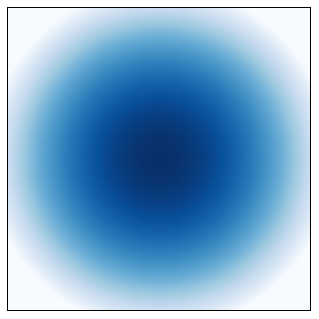}};
        \node (P_text) at (image6.north) {\footnotesize $\color{tikzblue} P$};
        \node (image5) at (5,-6.2) {\includegraphics[scale = 0.2]{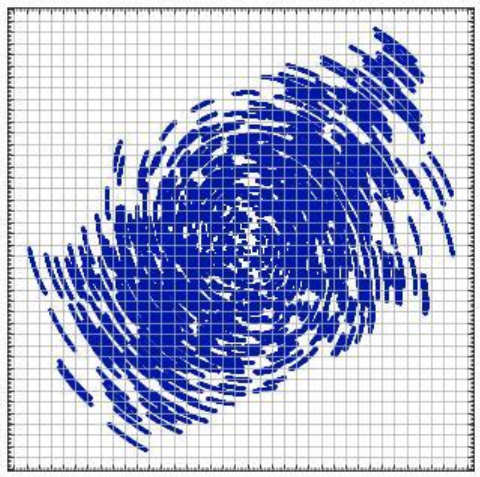}};
        \node (S_text) at (image5.north) {\footnotesize $\color{tikzblue} S$};
        \draw[->, shorten <=1pt, shorten >= 25pt] (box2.south) -- (A.north); 
        \draw[->, shorten <=0pt, shorten >= 25pt] (box2.south) -- (V.north); 
        \draw[->, shorten <=0pt, shorten >= 8pt] (box2.south) -- (N.north); 
        \draw[->] (box1.south) -- (image1.north);
        \draw[->] (box1.south) -- (image2.north);
        \draw[->, shorten <=1pt, shorten >= 8pt] (title1.south) -- (box4.north); 
        \draw[->, shorten <=2pt, shorten >= 1pt] (5.2,-2.52) -- (A.east);
         \draw[->, shorten <=1pt, shorten >= 8pt] (fourier.west) to [bend left=10] (A.south);
         \draw [->, shorten <=1pt, shorten >= 8pt] (image5.west) to [bend left=20] (A.south); 
         \draw [->, shorten <=1pt, shorten >= 8pt] ($(image6.west) + (0,0.5) $) to [bend left=30] (A.south); 
        \draw[->, shorten <=1pt, shorten >= 8pt] (V.south) to [bend right]  (box5.north); 
        \draw[->, shorten <=1pt, shorten >= 8pt] (A.south) to [bend left]  (box5.north);
        \draw[->,shorten <=1pt, shorten >= 4pt] (N.south) -- (box5.north);
    \end{tikzpicture}
    \caption{\ Diagram of the methodology used in this work to sample from the posterior using a score-based model as a prior.}
    \label{fig:method}
\end{figure*}

\subsection{Sampling from the posterior} 
Anderson's reverse-time formula Eq. \eqref{eq:reverse_sde} can be used to sample from the posterior distribution, simply by substituting the annealed prior $p_t (\mathbf{x}_t)$ by the annealed posterior $p_t(\mathbf{x}_t\mid \mathcal{V})$. The reverse SDE therefore becomes 
\begin{align}\label{eq:reverse_sde_posterior}
    d \mathbf{x}_t = \left[-\frac{1}{2}\beta(t)\mathbf{x}_t - g^{2}(t) \grad_{\mathbf{x}_t}\log p_t(\mathbf{x}_t \mid \vis) \right]\bar{dt} + g(t) d \bar{\mathbf{w}}_t \, .
\end{align}
To sample the posterior, one could train a SBM conditioned on the observed visibilities $\vis$ and substitute it in the reverse SDE. However, doing so has a few shortcomings. First, this approach leaves the physics of the measurement process and the noise model to be learned by the neural network. Although this might be acceptable if the likelihood is not readily available, it is not ideal for radio interferometry, where the physics and noise characteristics are well-understood. Moreover, it would require training a score-based model on a very high-dimensional space (as visibilities massively outnumber the pixels in a gridded image) to ensure that the neural network has coverage over a wide range of the possible pairs $(\mathbf{x}, \vis)$. As such, it is not guaranteed that a neural network trained in this way will generalize well on real-world data.
One way to bypass these issues is to instead write the posterior's score $\grad_{\mathbf{x}_t} \log p_t(\mathbf{x}_t \mid \vis)$ as the sum of the prior's score $\grad_{\mathbf{x}_t} \log p_t(\mathbf{x}_t)$ and the likelihood's score $\grad_{\mathbf{x}_t} \log p_t(\vis \mid \mathbf{x}_t)$ i.e. 
\begin{align}\label{eq:score_bayes}
    \grad_{\mathbf{x}_t} \log p_t(\mathbf{x}_t \mid \vis) = \grad_{\mathbf{x}_t} \log p_t(\mathbf{x}_t) +  \grad_{\mathbf{x}_t} \log p_t(\vis \mid \mathbf{x}_t) \, ,
\end{align}
where we can substitute the score of the prior by a trained SBM $\grad_{\mathbf{x}_t} \log p_t(\mathbf{x}_t) \approx \mathbf{s}_{\theta}(\mathbf{x}_t, t)$ and where the annealed likelihood $p_t(\vis \mid \mathbf{x}_t)$ can be analytically approximated. Since the neural network is now completely independent of the likelihood, the SBM acts as a versatile \textit{plug-and-play prior} which can be exploited to solve a variety of inverse problems. For example, the SBM framework used in this work has already been used to address radically different inverse problems such as source characterization in strong gravitational lenses \citep{Adam2022, Karchev2022}, PSF deconvolution with the Hubble Space Telescope \citep{Adam2023}, magnetic resonance imaging reconstruction \citep{chung2022_mri, Song2022}, black hole event horizon imaging \citep{Feng2023a, Feng2023b}. This plug-and-play aspect of the approach is agnostic to the radio interferometric data and can be used for any measurements as long as we are able to properly calculate the likelihood. We now derive an analytical estimate for the annealed likelihood using the \textit{Convolved Likelihood Approximation} \citep{Remy2022,Adam2022} (CLA), the core approximation of this work. 
\newpage

\subsection{The Convolved Likelihood Approximation}\label{sec:cla}
The annealed likelihood's score $\nabla_{\mathbf{x}_t} \log p_{t} (\vis\mid\mathbf{x}_t)$ is typically an intractable quantity \citep{Chung2022,Feng2023a,Feng2023b}. The likelihood of an observation (in this case, the visibilities $\vis$) given a noisy sample $\mathbf{x}_t$ (here, an image $\mathbf{x}_0$ passed to Eq. \eqref{eq:noising}) can be written using Bayes' theorem as 
\begin{align}
p_t(\vis \mid \mathbf{x}_t ) &= \int d \mathbf{x}_0 p_0(\vis \mid \mathbf{x}_0) p(\mathbf{x}_0 \mid \mathbf{x}_t)
\end{align}
This is intractable because $p(\mathbf{x}_0 \mid \mathbf{x}_t)$ is unknown. We can use Bayes theorem to write this equation as
\begin{align}
p_t(\vis \mid \mathbf{x}_t ) &= \int d \mathbf{x}_0 p_0(\vis \mid \mathbf{x}_0) p(\mathbf{x}_t \mid \mathbf{x}_0) \frac{p_0(\mathbf{x}_0)}{p_t(\mathbf{x}_t)}
\end{align}
but the ratio of the two marginal probabilities is also unknown. The CLA involves in treating the ratio $p(\mathbf{x}_0)/p(\mathbf{x}_t)$ as a constant or setting it to 1,
\begin{align}
p_t(\vis \mid \mathbf{x}_t ) &\approx \int d \mathbf{x}_0 p_0(\vis \mid \mathbf{x}_0) p(\mathbf{x}_t \mid \mathbf{x}_0) \, . \label{eq:cla}
\end{align}
The argument in favor of this simplification is that at low temperature ($t \sim 0$), the prior and the convolved prior are roughly equal, i.e. $p(\mathbf{x}_t) \approx p(\mathbf{x}_0)$. Hence, this approximation converges to the true likelihood at low temperatures. At high temperatures ($t \sim 1$), this approximation is much worse unless $p(\mathbf{x}_0)$ can be treated as a constant over the region where $p(\mathbf{y} \mid \mathbf{x}_0)$ contributes to the integral. In other words, the convolved likelihood approximation is more accurate when the likelihood is informative (narrow). Since the dynamics of the stochastic process are approximated at higher temperatures, the CLA only enables us to sample a conditional distribution close to the posterior. Although, other studies explore other possible approaches to sampling the posterior \citep{Feng2023a, Feng2023b, feng2024eventhorizonscaleimagingm87different, wu2024principledprobabilisticimagingusing}, we show through calibration tests on simulations Sec. \ref{sec:sims} that the CLA is appropriate for the specific problem considered in this work.

We now rewrite Eq. \eqref{eq:vis} for all temperatures (using our vectorized representation). We first multiply both the visibilities $\vis$ and the noise realization $\boldsymbol{\eta}$ by $\mu(t)$ in order to separate the likelihood from additional terms and define the residuals $\boldsymbol{\eta}_t$ at a time $t$
\begin{align}
    \nonumber
    \boldsymbol{\eta}_t &= \mu(t)\vis - A\mathbf{x}_t  
    \\
    &= \mu(t)\boldsymbol{\eta} - A\sigma(t)\mathbf{z}\, \label{eq:res_t} ,
\end{align}
where we used the reparametrization Eq. \eqref{eq:noising} at the second line. Since $\mu(0) = 1$ for the SDEs used in this work, Eq. \eqref{eq:res_t} converges to the residuals $\boldsymbol{\eta}$ as $t\rightarrow 0$. We can therefore rewrite Eq. \eqref{eq:cla} in terms of the variables $\boldsymbol{\eta}$ and $\boldsymbol{\eta}_t$
\begin{align}\label{eq:eta_cla}
    p_t (\boldsymbol{\eta}_t) &=  p_t(\vis\mid \mathbf{x}_t)\\
    &= \int d \boldsymbol{\eta}\, p(\boldsymbol{\eta}) p(\boldsymbol{\eta}_t \mid \boldsymbol{\eta})
\end{align}
where the density function $p(\boldsymbol{\eta}_t \mid \boldsymbol{\eta})$ is obtained by applying the change of variables formula with the perturbation kernel $p(\mathbf{x}_t \mid \mathbf{x}_0)$. We finally evaluate the convolution implied by the sum of random variables Eq. \eqref{eq:res_t} (note that each term follows a Gaussian distribution) and arrive to the expression of the convolved likelihood 
\begin{equation}\label{eq:convolved_likelihood}
    p_t(\vis \mid \mathbf{x}_t) \approx \mathcal{N}
    \left(
    \mu(t)\vis \mid A\mathbf{x}_t, \mu^2(t)\Sigma + 
    \sigma^2(t)\Gamma 
    \right)\, ,
\end{equation}
where $\Gamma \equiv AA^{T}$. For the physical model used in this work, we approximate the covariance matrix as diagonal (we derive in Appendix \ref{app:cla} that the covariance matrix can be approximated as $\Gamma_{jk}\approx \frac{1}{2}\left(\delta_{jk} + \delta_{j0}\delta_{k0}\right)$). As long as the forward model is linear, the hereby expression can be used along a trained SBM $\mathbf{s}_{\theta}(\mathbf{x}_t,t)$ in Eq. \eqref{eq:reverse_sde_posterior} to effectively sample the posterior distribution. We summarize the proposed methodology in Figure \ref{fig:method}.  

\section{Data and physical model}\label{sec:data_and_phys}
\subsection{Training sets for the score-based priors}\label{sec:sbms}
The PROBES dataset is a compendium of high-quality local late-type galaxies \citep{Stone2019,Stone2021} that we leverage as a prior for our reconstructions. These galaxies, used in previous studies to train diffusion models \citep{Smith2022,Adam2022}, have resolved structures of spiral arms, bulges, and disks, which can vaguely resemble the structure of protoplanetary disks, in comparison to terrestrial objects and scenes or many other astronomical images.

The SKIRT TNG \citep{Bottrell2023} dataset is a large public collection of images spanning $0.3$-$5$ microns made by applying dust radiative transfer post-processing \citep{Camps2020} to galaxies from the TNG cosmological magneto-hydrodynamical simulations 
\citep{Nelson2019}. The SKIRT TNG dataset offers us an opportunity to compare prior distributions based on the same type of objects --- galaxies --- but with different underlying assumptions about the physics that govern the formation of their structure \citep{Weinberger2017,Pillepich2018}, which in the case of SKIRT TNG is inherited from simulations instead of observations. In this work, we use the $z$ band in both datasets to train the score-based priors on $256 \times 256$ images.  

\subsection{DSHARP survey and data preprocessing}
The Disk Substructures at High Angular Resolution Project (DSHARP) \citep{dsharp1_2018, dsharp2_2018, dsharp3_2018, dsharp4_2018, dsharp5_2018, dsharp6_2018, dsharp7_2018, dsharp8_2018, dsharp9_2018, dsharp10_2018} is a recent public survey conducted during ALMA's Cycle 4 to capture high-resolution ($\sim0.035''$) observations of the continuum emission from 20 nearby ($\sim140$ pc) protoplanetary disks around 240 GHz (Band 6). Protoplanetary disks play a crucial role in planetary system formation, generating significant interest in understanding the connection between disk properties and the characteristics of the corresponding exoplanet population. However, existing theoretical frameworks have limitations in explaining these relationships \citep{Bae2023}. The primary goal of DSHARP is to observe protoplanetary disk substructures—such as rings, gaps, spirals, and crescents—to provide the observational data necessary to improve these theoretical models. Accurate imaging of these disk features is therefore a critical step towards a better understanding of the physical processes occurring during the early stages of planetary system evolution.

We process the calibrated visibilities released by the DSHARP survey using the \texttt{visread} package created by the MPoL team \citep{Zawadzki2023}. For each spectral window, we average the data across the different polarizations and conserve the non-flagged data. To combine spectral windows for continuum imaging, we assume a flat spectral index. In order to perform inference over regular grid models, we bin the visibilities using our implementation of a sinc convolutional gridding function, which was chosen for its uniform effect in image space. We estimate the diagonal part of the covariance matrix, $\Sigma \in \mathbb{R}^{2m \times 2m}$, by computing a weighted standard deviation of the real and imaginary parts of the observed visibilities within each bin. If a bin contains a small number of visibilities, we simultaneously expand the bin and the sinc window function until each cell has at least 5 data points. 

The proposed approach does not require any preprocessing techniques involving visibility weighting schemes \citep{Briggs_thesis1995} typically used in the field to reduce the strength of the artefacts in the dirty image.

\subsection{Physical model}
In a statistical inference setting, the physical (forward) model is a mapping between the parameters of interest and simulated data. In the context of this problem, it describes the physical relationship between a sky brightness $I(l,m)$ that we wish to infer, and a visibility function $V(u,v,w)$ measured by a radio interferometer. In an ideal case where noise is absent and where electromagnetic waves can travel freely through space, a radio interferometer would measure 
\begin{align}\label{eq:van_theorem}
    V(u,v&,w) = S(u,v,w)\iint P(l,m)I(l,m) \nonumber \\
    \exp&\Bigl\{ -i2\pi \left[ul+vm+w\left(\sqrt{1-(l^2 +m^2) }-1\right)\right]\Bigr\} \nonumber \\
    \Bigl[1-&\left(l^2 + m^2\right) \Bigr]^{-1/2} dl\,dm .
\end{align}

The $(u,v,w)$ components form a right-handed coordinate system allowing one to measure baselines, the distances (in units of wavelengths) between pairs of antennae in a radio interferometer. The $(l,m)$ components are direction cosines measured with respect to the $u$ and $v$ axis. $S(u,v,w)$ is known as the sampling function which represents the regions of the spatial frequency space that are effectively probed by the interferometric array. It is typically modeled as a binary function, taking the value $1$ where data is measured (or sampled), and $0$ elsewhere. The sampling function is related to the synthesized (dirty) beam $B(l,m)$ through the Fourier operator $F$, such that $B(l,m)= F^{-1}S$. This quantity is commonly used in other imaging algorithms to compute the dirty image $\tilde{I}$ as a function of the sky brightness (i.e. $\tilde{I} = F^{-1}(SV) = I*B$ where $*$ denotes the convolution operation) therefore reframing the radio interferometric imaging problem as a deconvolution task. Finally, $P(l,m)$ is the primary beam encoding the response function of each antenna. It can be assumed to be the same for each antenna of the array and modeled as a Gaussian whose full-width half maximum (FWHM) is defined as the diameter of the field of view (in sky brightness space) of the radio interferometer. 

We now use several approximations to represent Eq. \eqref{eq:van_theorem}. We employ the coplanar baseline assumption by setting $w=0$ and assume a narrow field for the sky brightness, i.e. $l^2 + m^2 \ll 1$. The visibility and the sampling function therefore become functions of $u$ and $v$ only. We can simplify Eq. \eqref{eq:van_theorem} and write the visibility function using the two-dimensional Fourier transform
\begin{align}
    &V(u,v) \nonumber \\
    &\approx S(u,v)\iint \, P(l,m)I(l,m)e^{-i2\pi \left(ul+vm\right)} dl \, dm \, \\
    &= S(u,v)F\left[P(l,m)I(l,m)\right] \label{eq:forward_model_continuous}\, .
\end{align}

Eq. \eqref{eq:forward_model_continuous} can be discretized by defining the sky brightness and the visibility function on a regular grid with constant pixel size. As such, the forward model maps the pixelated sky emission $\mathbf{x}$ to the gridded visibilities $\mathcal{V}$ via the following complex matrix 
\begin{align}\label{eq:forward_model}
    \tilde{A} \equiv S\mathcal{F}P \, ,
\end{align} 
where $\mathcal{F} \in \mathbb{C}^{n\times n}$ is the dense, unitary 2D discrete Fourier operator, $S\in [0,1]^{m\times n}$ is a 2D binary mask representing the sampling function and $P\in \mathbb{R}^{n \times n}$ is the pixelated primary beam. We model $\mathcal{F}$ with a Fast Fourier Transform (FFT) \citep{CooleyFFT1965} and assume a circular Gaussian antenna response. In order to make our likelihoods as informative as possible, we add a padding operation in our forward model which gives us the flexibility to increase our resolution in Fourier space without modeling the entire field of view in image space. In practice, we infer the quantity $P\mathbf{x}$ and then multiply our posterior samples by $P^{-1}$ (since $P$ is a diagonal matrix) to obtain the sky brightness $\mathbf{x}$ since including the primary beam in the forward model complicates the approximation of the matrix $\Gamma$ (see Figure \ref{fig:cov_gamma} in Appendix \ref{app:gamma_approx}).

Although the true complexity of a radio interferometer's measurement process is represented by a nonlinear forward model, the Radio Interferometric Measurement Equation \citep{Noordam1996, Smirnov2011}, Eq. \eqref{eq:forward_model} provides us with a convenient linear forward model to perform posterior sampling with score-based models.
Creating a physically more realistic model that would allow one to perform polarization or multi-frequency imaging is outside of the scope of this work.

\begin{figure*}[!ht]
    \vspace{-0.75cm}
    \centering
    \begin{tikzpicture}
        \centering
        \node (image1) at (0,0){\includegraphics[scale = 0.35]{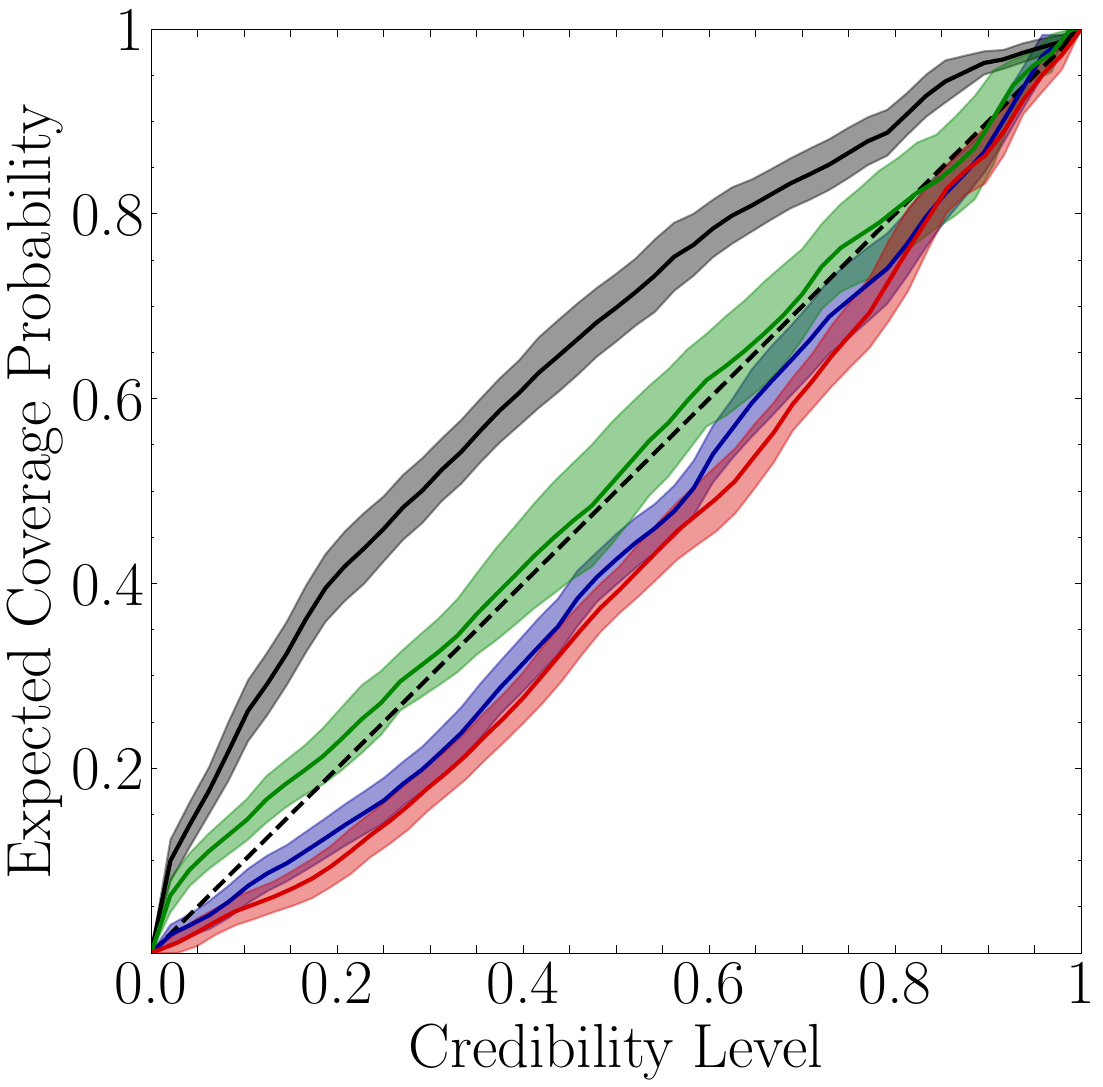}}; 
        \node (image2) at (8,0){\includegraphics[scale = 0.35]{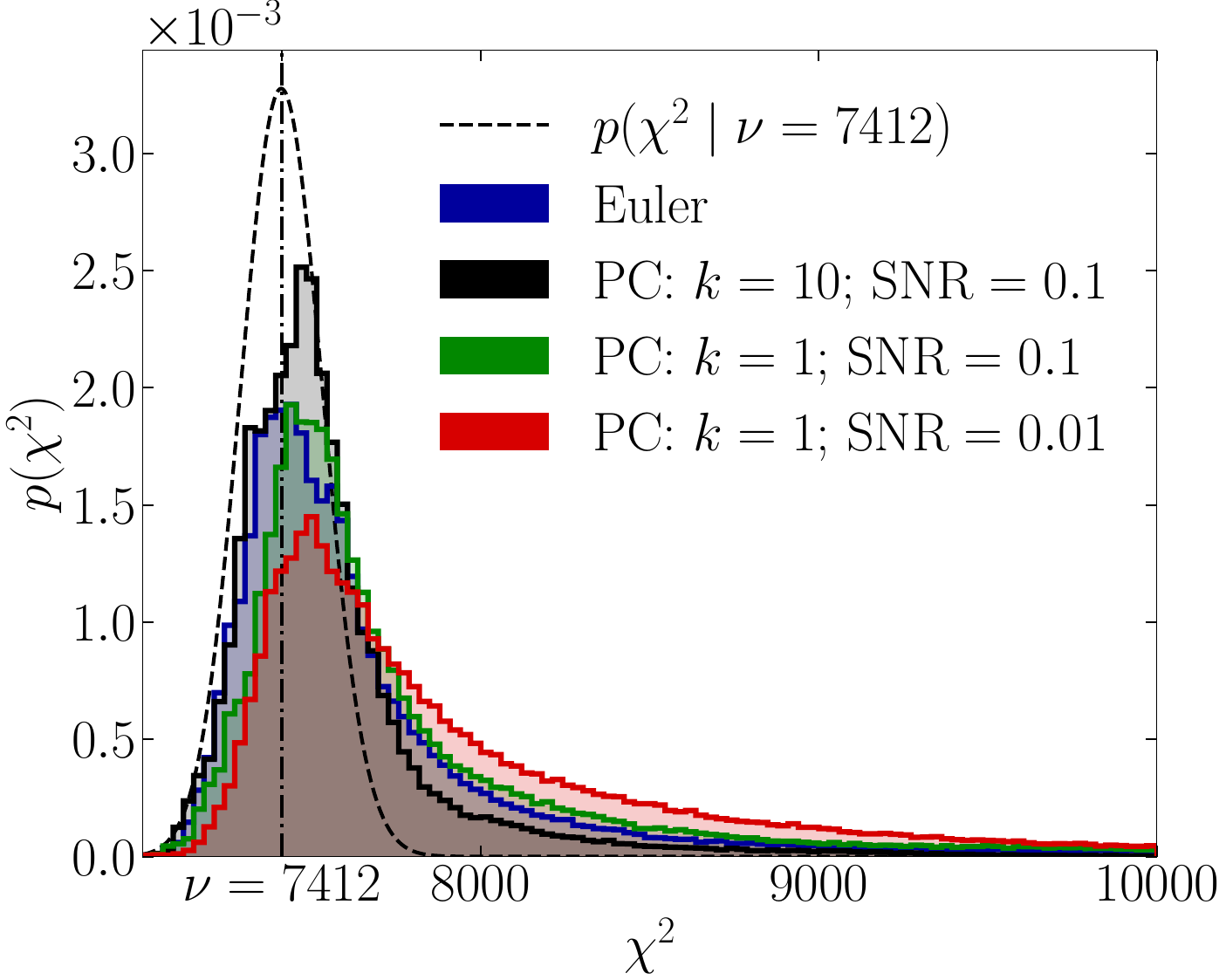}};
    \end{tikzpicture}
    \vspace{-0.3cm}
    \caption{Statistical tests results of posterior samples obtained via the Euler sampler and various combinations of corrector steps and SNR for the PC sampler with the proposed approach using a score model trained under VP SDE. Each curve is color-coded consistently across both figures. Left: TARP coverage test. The shaded regions show a 99.7\% confidence interval over multiple TARP tests computed with bootstrapping and the plain curve correspond to the mean. The dashed line represents the ideal case where the posterior estimator is calibrated. Right: Histograms of the samples obtained by computing the $\chi^2$ of our posterior samples.}
    \label{fig:tarp_chi}
\end{figure*}

\begin{figure*}[!ht]
    \centering
    \vspace{-0.4cm}
    \begin{tikzpicture}
        \centering
        \node (image1) at (0,0){\includegraphics[scale = 0.35]{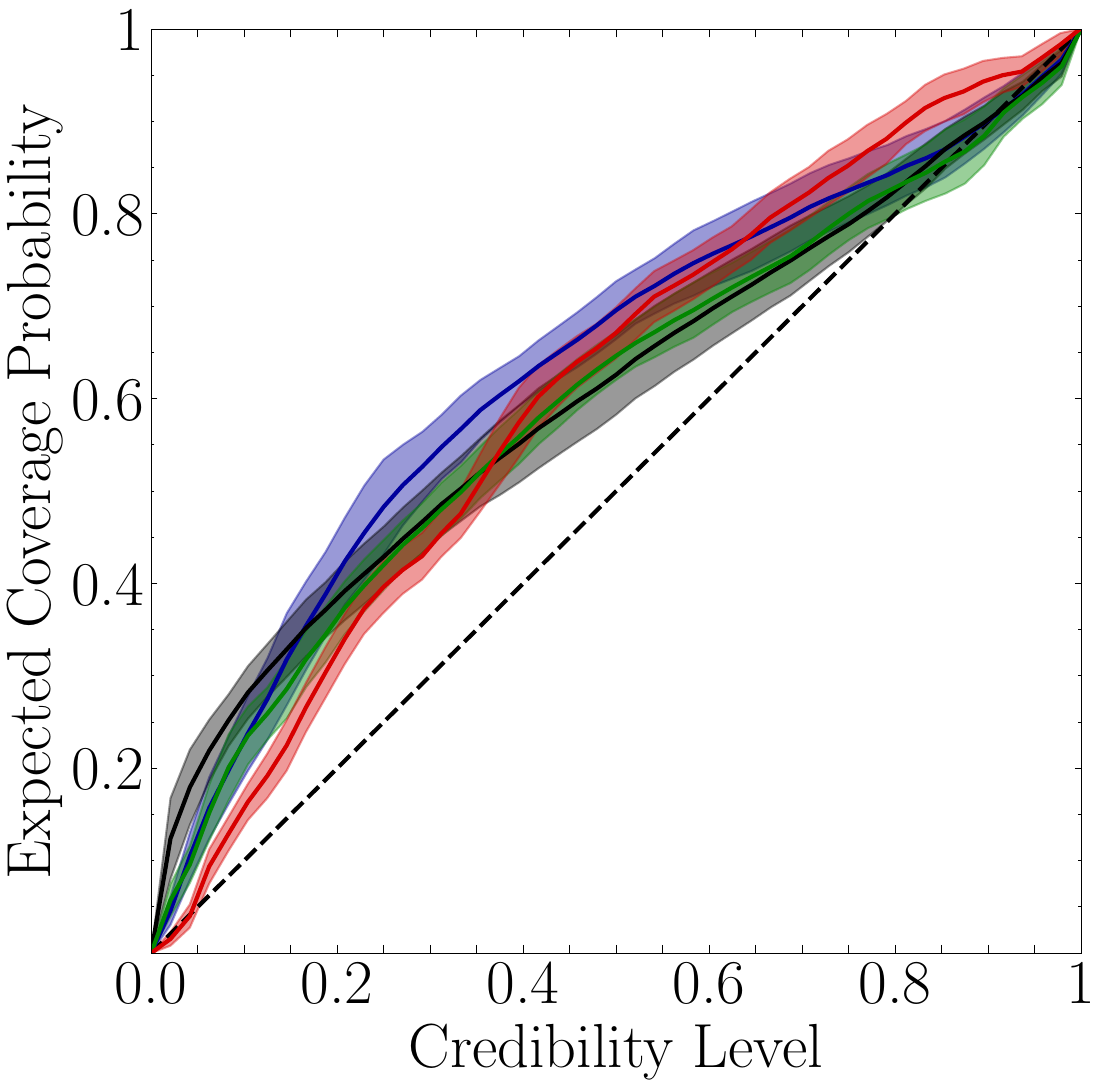}}; 
        \node (image2) at (8,0){\includegraphics[scale = 0.35]{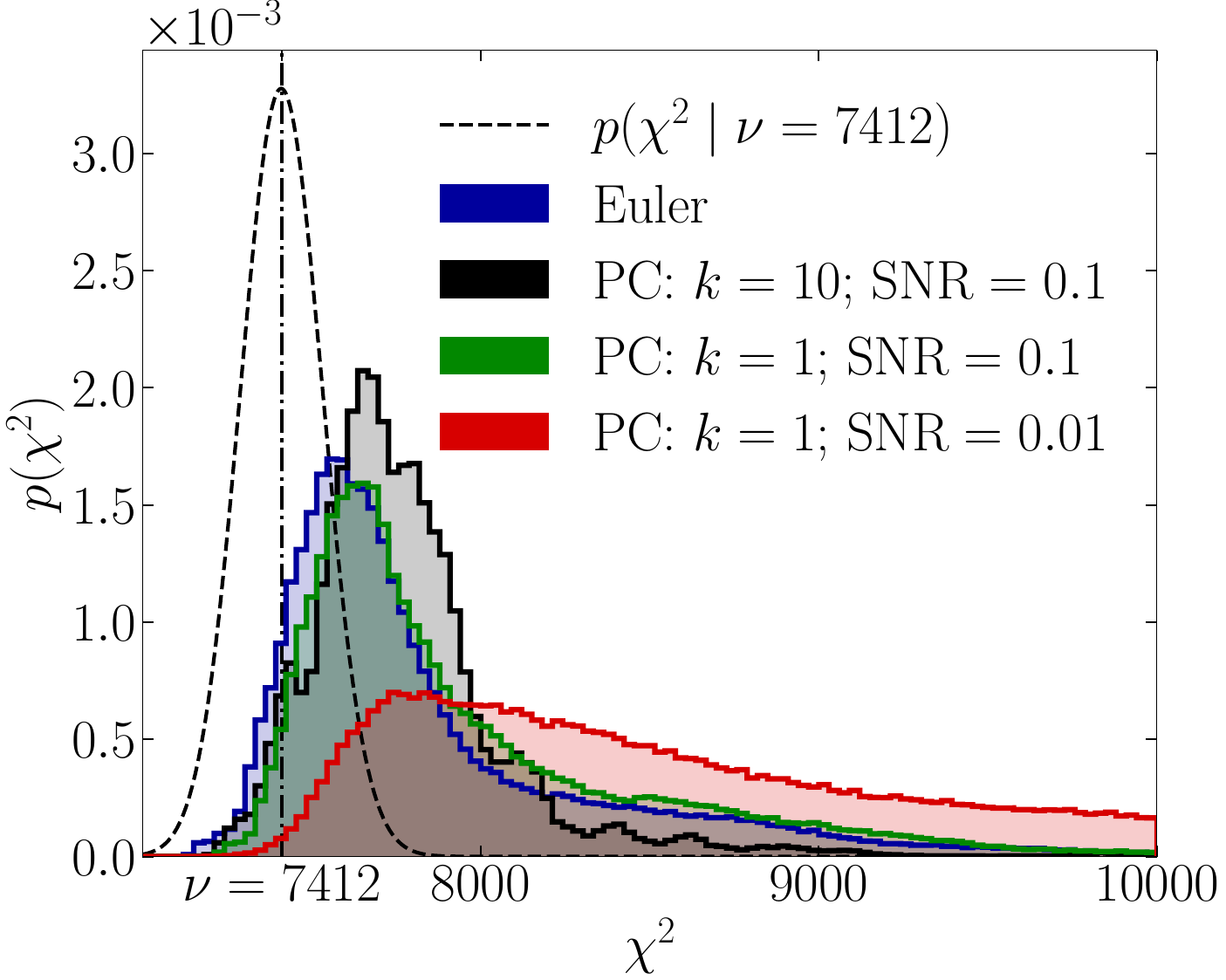}};
    \end{tikzpicture}
    \vspace{-0.3cm}
    \caption{Same statistical tests as in Figure \ref{fig:tarp_chi}, but for a score model trained under VE SDE.}
    \label{fig:tarp_chi2_ve}
\end{figure*}

\section{Tests on simulations}\label{sec:sims}
To explore the potential limitations of this approach, we test our imaging algorithm within a controlled framework using simulations. We train two separate SBMs: one for the SKIRT dataset and another for the PROBES dataset, both downsampled to $64\times 64$ pixels. We generate synthetic visibilities observations on a $256\times 256$ pixel grid by forward modelling samples drawn from these score-based priors and adding isotropic Gaussian noise to the result. For the sampling function, we choose the Fourier coverage for the specific interferometer layout used to observe HT Lup, one of the DSHARP protoplanetary disks. We assess the coverage of the posterior using Tests of Accuracy with Random Points \citep[TARP;][]{Lemos2023}, a method to estimate the calibration of a posterior estimator with only access to its samples and the associated ground-truth. We generate 500 synthetic observations and 500 posterior samples per observation (for a total of 250 000 posterior samples) in $\sim$ 1.5 hrs of wall-time using 500 A100 GPUs running in parallel for a total of $\sim 0.09$ GPU-years. This posterior sampling procedure is performed for both score-based priors and for the Euler sampler the PC sampler. For the Euler sampler, we take 4000 steps. For the PC sampler, we test multiple combinations of corrector steps $k$ and SNR values, with each combination involving 4000 predictor steps. The total amount of computation for this test for the two score-based priors and the various sampler parameters tested is therefore $0.72$ A100 GPU-years. We show the results of the TARP test and of a $\chi^2$ goodness-of-fit test for the posterior samples obtained with the VP SKIRT prior in these different settings in Figure \ref{fig:tarp_chi}. Posterior samples obtained through the Euler sampler deviate significantly from the ideal case which indicates a bias in the outlined approach. However, we empirically show that with the VP SDE, and the right set of parameters, this bias can be alleviated using the PC sampler. For the predicted visibilities, we note that calibrated posterior samples do not necessarily achieve the best residuals. We believe this effect is due to the limits of the CLA and that a better approach would be needed to be able to achieve both posterior calibration and accurate representation of the likelihood in the posterior samples. We show in Figure \ref{fig:tarp_chi2_ve} statistical results for the VE SDE in the same settings. However, we do not find PC parameters yielding calibrated posterior samples.

These TARP tests provide empirical evidence indicating that the CLA samples a distribution close to the unbiased posterior. Conducting a thorough analysis of how the calibration of the posterior might be affected by various factors (e.g., changes in the forward model or the dimensionality of the images) is beyond the scope of this work. 
This is an important avenue for future investigation to achieve unbiased uncertainty quantification with the proposed imaging algorithm. We explore both samplers for the application of this approach to ALMA data.

\begin{figure*}[!p] 
    \centering
    \hspace{-1cm} 
    \includegraphics[height=\textheight]{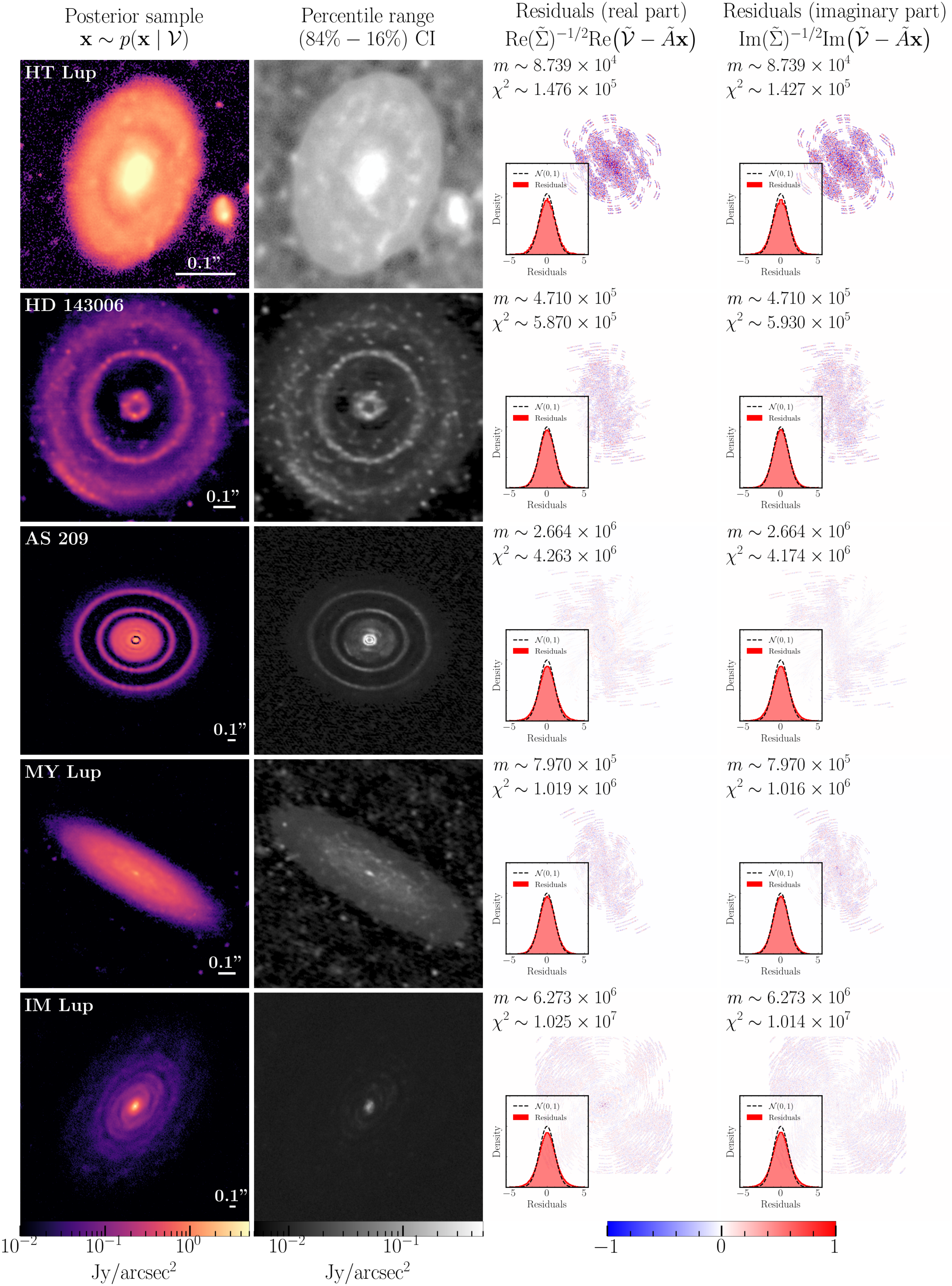}
\end{figure*}

\begin{figure*}[!p]
    \centering
    \hspace{-1cm}
    \includegraphics[height=\textheight]{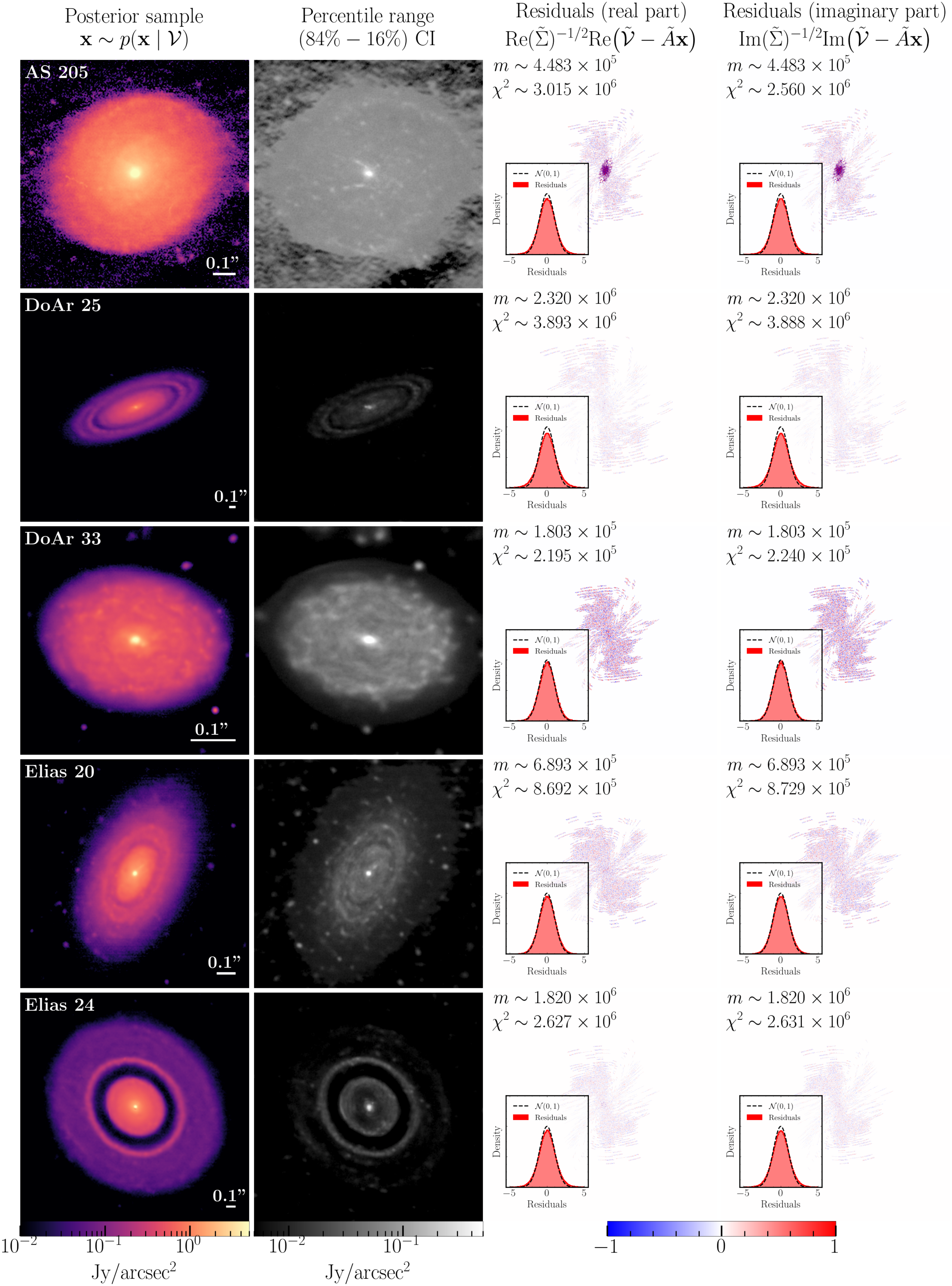}
\end{figure*}

\begin{figure*}[!p]
    \centering
    \hspace{-1cm}\includegraphics[height=\textheight]{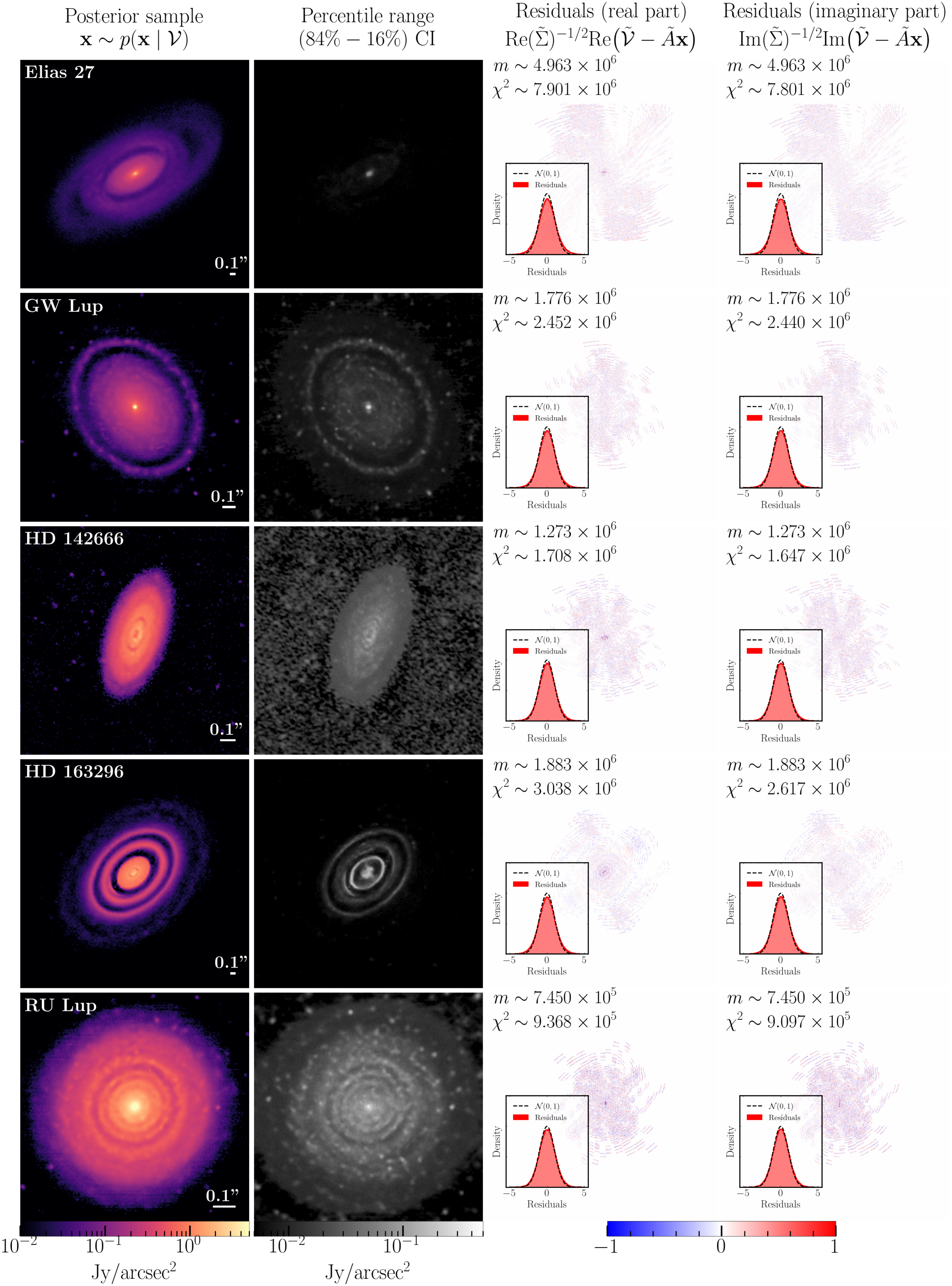}
\end{figure*}

\begin{figure*}[!htbp]
    \centering
    \hspace{-1cm}\includegraphics[height=0.91\textheight]{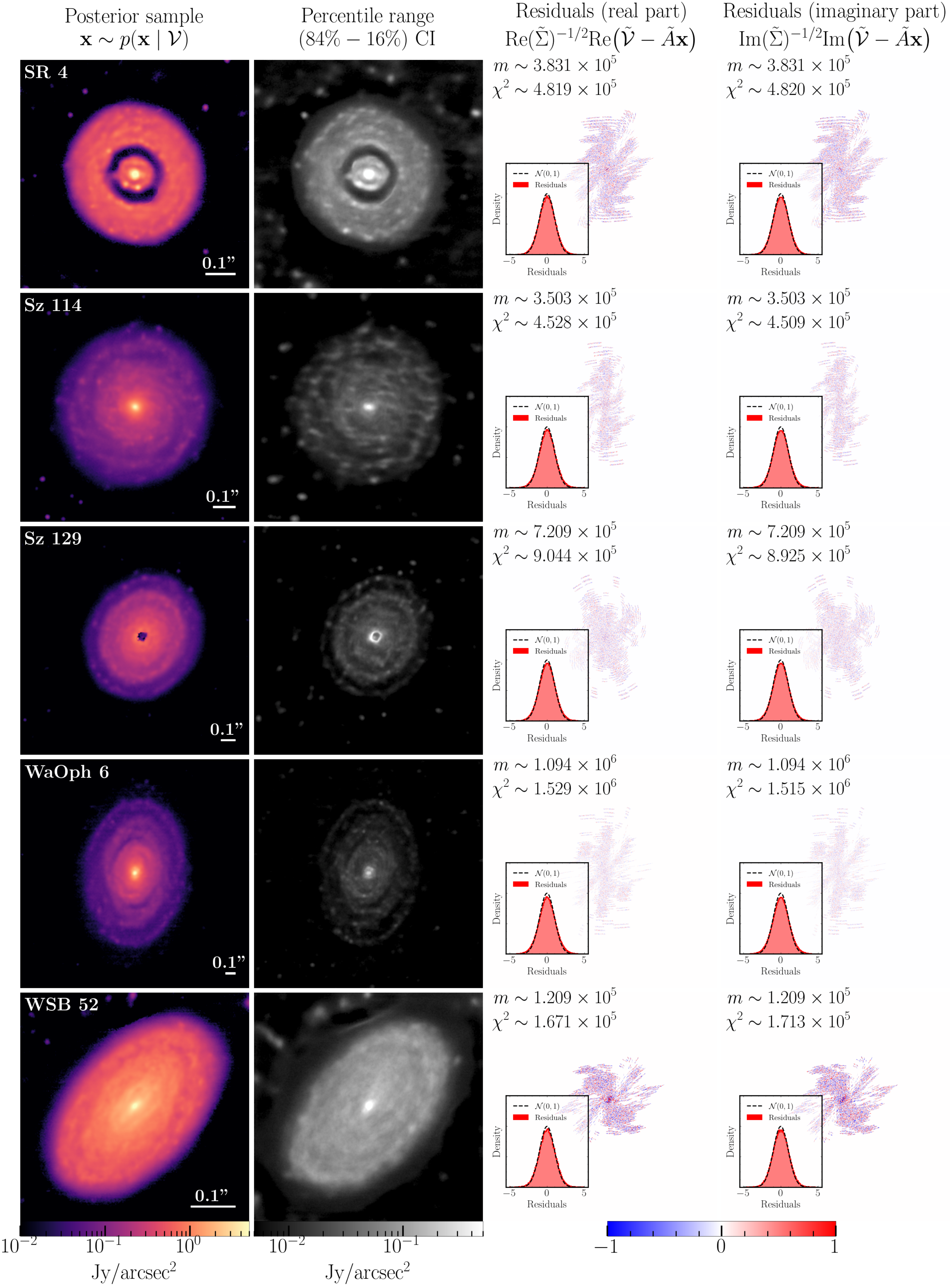}
    \caption{The proposed approach applied on the protoplanetary disks from the DSHARP survey using the VE PROBES score-based prior. From left to right: posterior sample, percentile range (pixel-wise) and residuals (real and imaginary parts) for each DSHARP protoplanetary disk imaged with IRIS using the VE PROBES prior. For each disk, we show on the posterior sample figure a scale bar for $0.1''$ on the bottom right corner.}
    \label{fig:all_disks}
\end{figure*}

\section{Application on DSHARP data}\label{sec:results} 
We now apply our approach to ALMA data with the DSHARP survey. To handle the prior misspecification, we make a few minor adjustments to our approach. First, since our training datasets described in section \ref{sec:data_and_phys} typically have smaller dynamic range compared to the sky brightness of the protoplanetary disks we wish to reconstruct, we normalize the visibilities (and thus the sky brightness due to the linearity of the Fourier operator) by a factor $s/\tilde{I}_\mathrm{max}$, where $\tilde{I}_\text{max}$ is the dirty image's brightest pixel and where $s$ is a scaling factor depending on the prior used and the protoplanetary disk. We show the effect of this factor on our results for the VP SKIRT prior Appendix \ref{app:scale_factor}. Secondly, some protoplanetary disks emit over a large spatial range, requiring a wider field of view to model the full sky emission. Because our score model operates on a constant pixel grid, we increase the pixel size for these protoplanetary disks. Consequently, the associated Fourier space is contracted, causing a fraction of the (high-frequency) observed visibilities to lie outside this field of view. We choose to ignore these visibilities in our modeling. While this approach is not ideal as it reduces the number of data points in an already ill-posed inverse problem, it serves as a temporary solution until the implementation of a more advanced forward model for multi-scale imaging \citep{Cornwell2008multiscale}.

With these few adjustments in place, we use 10 V100 GPUs to generate $250$ posterior samples (each sample being a $256 \times 256$ pixel image) in $\sim 2.5$ hours wall-time for each protoplanetary disk. Sampling is performed with 4000 steps of the discretized Euler-Maruyama SDE solver as PC sampler showed no significant improvements in our experiments (we show in Appendix \ref{app:sampler_comparison} a comparison between the two samplers for one of the DSHARP protoplanetary disks). We show in Fig. \ref{fig:all_disks} a posterior sample for each protoplanetary disk along with the real and imaginary part of the residuals in Fourier space and the associated $\chi^2$. The full 250 posterior samples for each disk are freely available on Zenodo\footnote{\href{https://zenodo.org/records/14454443}{https://zenodo.org/records/14454443}}. 
To quantify the goodness-of-fit, we present a histogram of the residuals and compare them to a standard Gaussian distribution (since the noise is expected to be highly Gaussian) and provide the $\chi^2$ values. Despite prior misspecification, we recover the key features (crescent, gaps, rings...) of all protoplanetary disks. Some of the residuals are higher in the central region, which might indicate a poor capability of our model in specific cases to predict the total flux. To address this issue, in future work we plan to train our SBMs with flux normalized data and to infer a total flux parameter along with the pixelated image we wish to reconstruct.

\begin{figure*}[!ht]
    \centering
    \includegraphics[width=\textwidth]{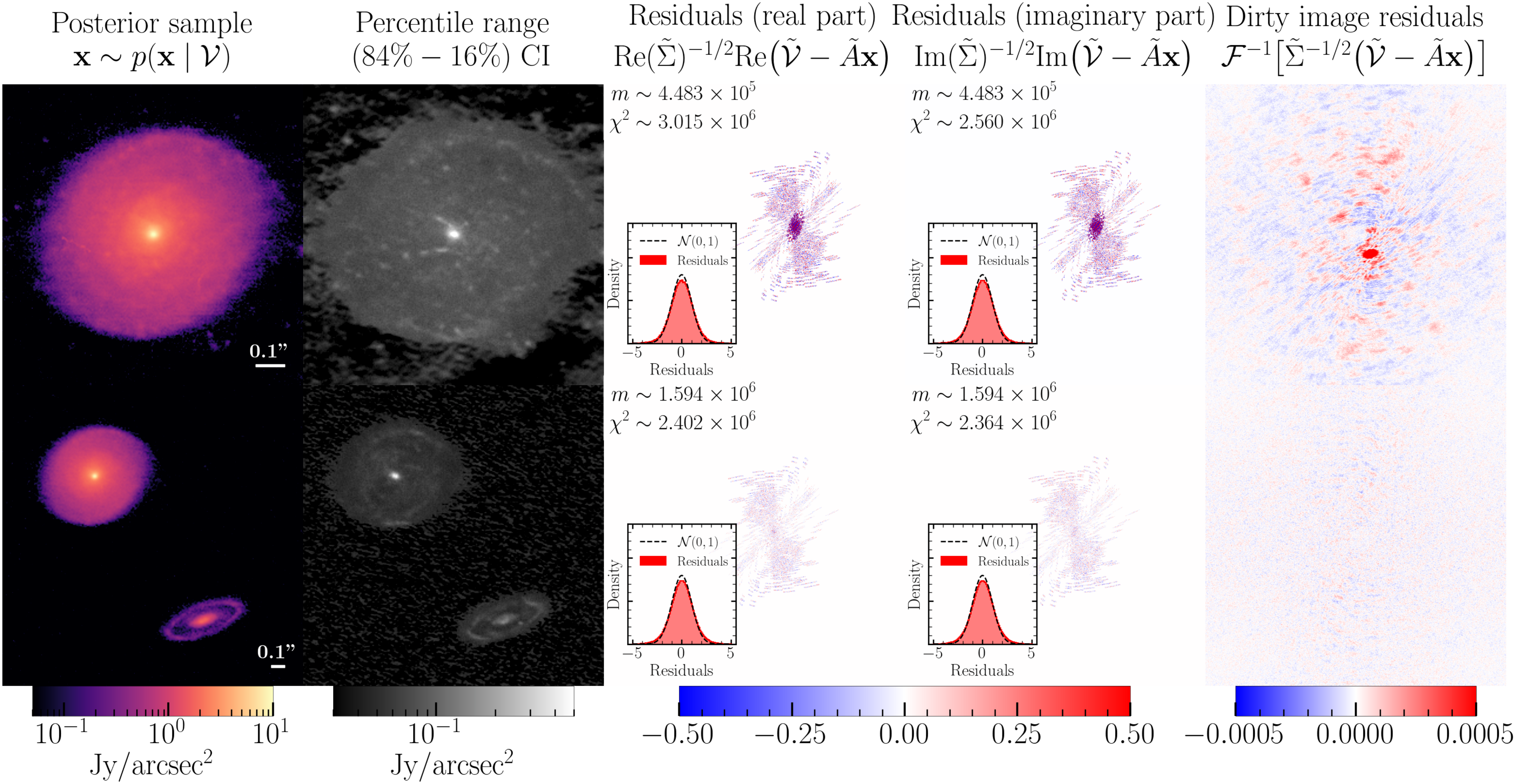}
    \caption{An example of companion modeling for the multiple system AS 205 with the VE PROBES prior using the imaging algorithm outlined in this work. The first row shows results of the inference on the AS 205-N disk only and the second row for the binary system AS 205. Each column shows from left to right a posterior sample, the pixel-wise percentile range, the residuals in Fourier space (real and imaginary parts) and the dirty image residuals. The dirty image residuals are shown for the padded image ($4097\times 4097$ pixels) to highlight the missing component of our modeling, AS 205-S first row. Residuals are computed using the posterior sample of each row.}
    \label{fig:as205_companion}
\end{figure*}

Note that the $\chi^2$ values shown in Figure \ref{fig:all_disks} are much higher than the degrees of freedom associated with the number of data points. The key factor in this difference is the fact that we only model a small region (of order of an arcsecond) of the field of view in image space with the remaining sky brightness assumed to be zero (over tens of arcsecond of the size of the primary beam). This is in general incorrect as there are usually other objects in the field of view, for example other protoplanetary disks, as they can be part of a multiple system \citep{dsharp4_2018}. In order to detect the presence of such companions, one can generate dirty images $\tilde{I}$ of the data or the residuals of the model as a diagnostic to identify such extended emissions. The sources themselves can be removed through peeling \citep{Williams2019}. We show an example Figure \ref{fig:as205_companion} of such companions and how modelling the principal components of the sky emission can significantly improve residuals and dirty image residuals for the AS205-N disk whose companion, AS205-S \citep{dsharp4_2018}, is close enough for both structures to be inferred simultaneously in a single image. We note that despite the prior misspecification, both structures are well recovered. This result aligns with previous works \citep{Adam2022, feng2024eventhorizonscaleimagingm87different,Rozet2024, Barco2024} showing the robustness of SBMs to high distributional shifts. In scenarios where the different structures are too distant to be inferred jointly while maintaining high angular resolution, we propose to first locate each structure using the dirty image and to attribute a score model to each of them. For the rest of our analysis, we will focus exclusively on protoplanetary disks without companions.

\subsection{Comparison with other imaging algorithms}
\begin{figure*}[!t]
\hspace{-0.5cm}
    \centering
    \includegraphics[width=\textwidth]{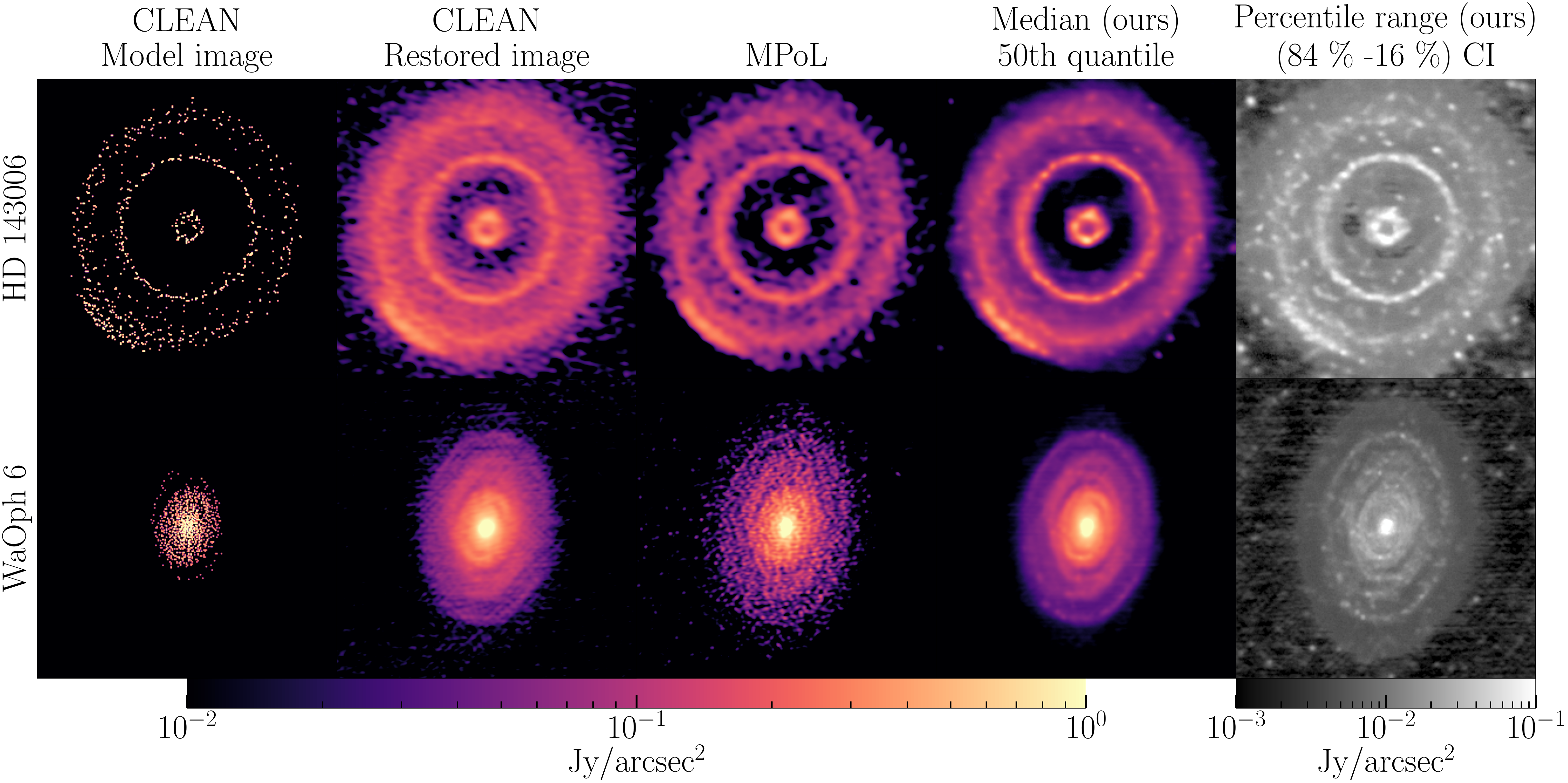} 
    \caption{Image reconstructions by CLEAN, MPoL, and our approach for the protoplanetary disks HD 143006 and WaOph 6. The two columns at right are pixel-wise statistics computed using the posterior samples obtained from our score-based approach using the PROBES prior (we also show posterior samples of these two disks in appendix \ref{app:method_comparison}). From left to right the statistics include the median and the 84\%-16\% percentile range. The MPoL image for WaOph 6 (bottom, second column from left) is not fully optimized and is shown purely for illustration.}
    \label{fig:method_comparison}
\end{figure*}

We now compare our approach to state-of-the art imaging algorithms, CLEAN and MPoL, in Figure \ref{fig:method_comparison} for HD 143006 and WaOph 6. The CLEAN images were obtained by running the \texttt{tclean} task from the Python package \texttt{casatasks}. For the weighting visibility scheme, we adopted the parameters specified in the original DSHARP work (see Table 4 in \cite{dsharp1_2018}). We did not, however, include a tapering function in order to have a better benchmark for comparison (since our algorithm does not include this tapering scheme). The model image represents CLEAN's first output after the iterative deconvolution: a collection of point sources (also known as CLEAN components). The model image is then convolved with a restoring beam to obtain the restored image, CLEAN's final output. For the MPoL image of HD 143006, we adopted the hyperparameters from the MPoL documentation’s tutorial, which were optimized for this disk; we then performed a limited hyperparameter search for the disk WaOph 6. Thus, the performance of MPoL on the disk WaOph 6 is not optimized and is shown only for illustrative purposes. To be able to compare algorithms in terms of dynamic range, we chose the same pixel size in image space across imaging algorithms. We recover all substructures present in MPoL and CLEAN and our results show competitive resolution and dynamic range performance. Note that, while the CLEAN deconvolution process results in persistent artifacts from the dirty image, such artifacts are absent from our reconstructed images.

While generating a large number of posterior samples with IRIS is computationally expensive, its key advantages lie in its ability to encode flexible priors over complex astrophysical objects and to recover statistical uncertainties through posterior samples. While CLEAN and MPoL remain significantly faster imaging algorithms, IRIS is a good alternative when accurately accounting for uncertainties is critical. However, this speed comparison assumes that the optimal hyperparameters required by alternative methods have already been identified. In situations where hyperparameters need to be determined manually for each individual source (e.g., for each disk in the DSHARP survey), our approach proves more versatile as, besides gridding, the only preprocessing step is the renormalization of the visibilities (which, as discussed above, is due to the brightness of sources in the prior being misspecified).

Another notable advantage of the proposed method is the complexity and realism of the prior used in the reconstruction. While it is misspecified, the galaxy structures captured by the score-based model allow the reconstruction of a wide variety of galactic disk features and their correlation across multiple scales. On the other hand, popular methods such as CLEAN assume a collection of point sources as prior, whereas other reconstruction methods such as MPoL impose parametric constraints as priors (e.g. maximum entropy or an analytical regularization on the gradient of the image pixels). In contexts where the complexity of the data is very high, as it is the case here, such parametric representations can be insufficient to capture all the information content of the data, which can result in biased inference. 

\begin{figure*}[!ht]
  \hspace{-0.75cm}
  \centering
  \begin{tikzpicture}
    \node at (0,0.) {\includegraphics[width=\textwidth]{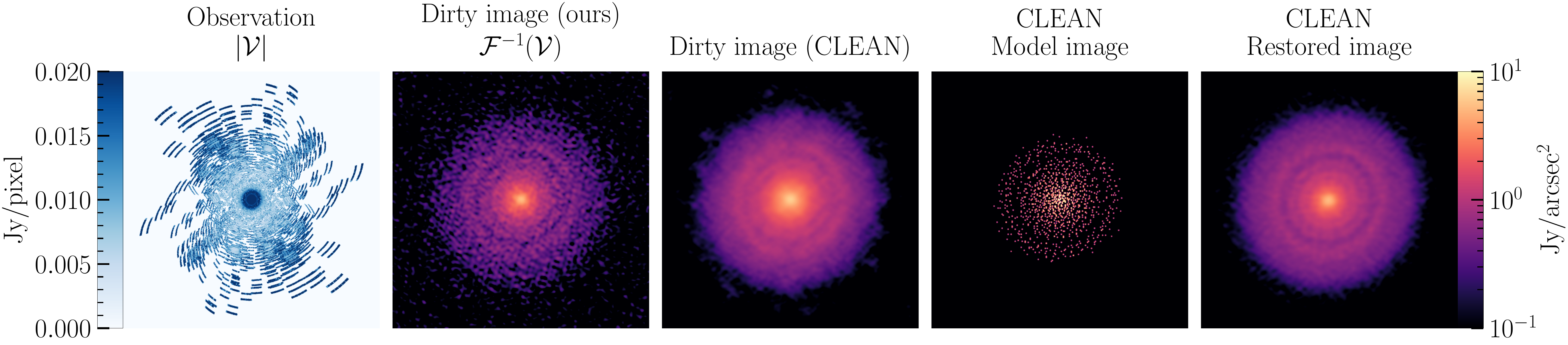}};
    \node at (0.,-8.) {\includegraphics[width = 1.05\textwidth]{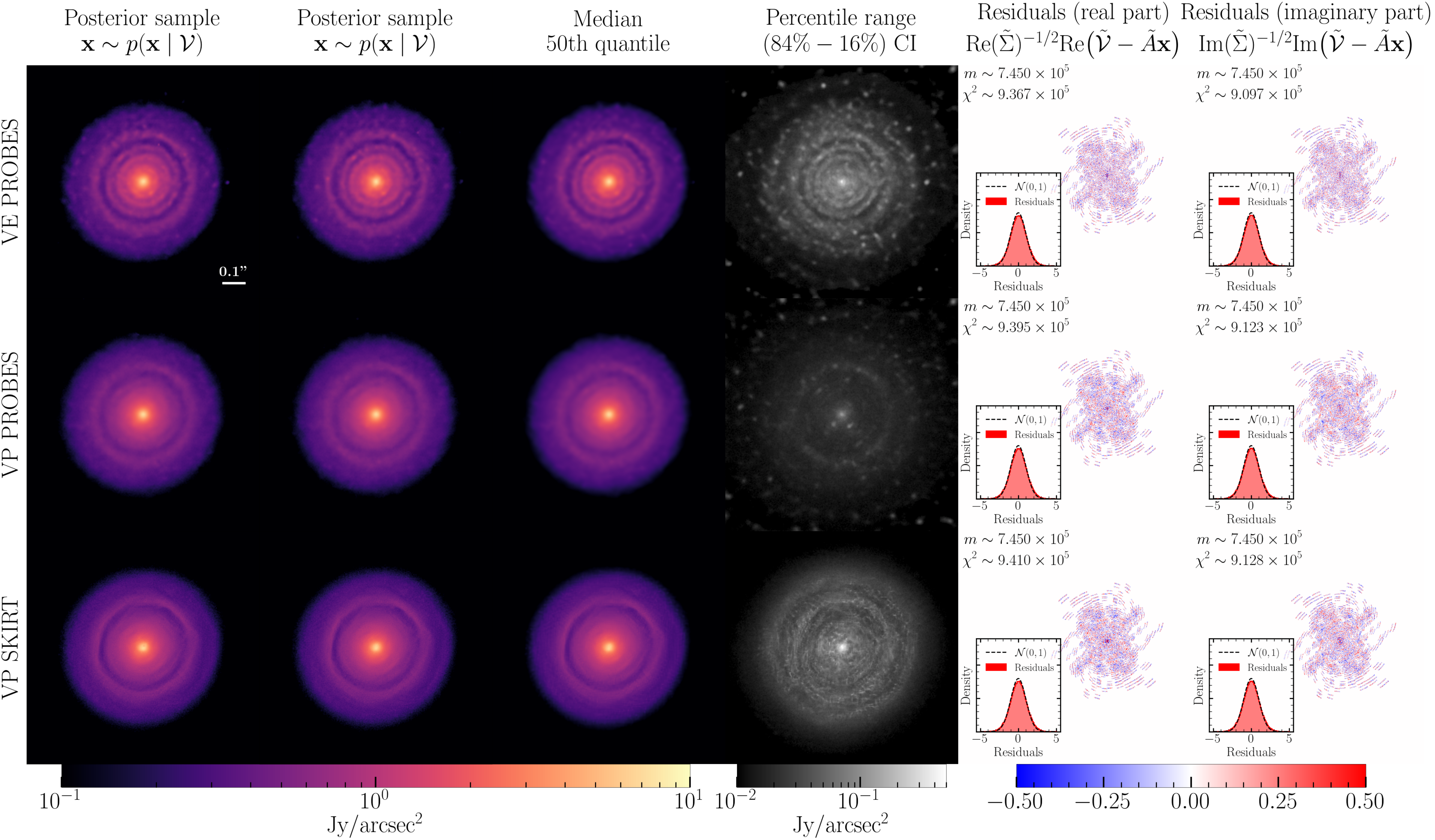}}; 
    \draw[-] (-9,-2.25) -- (9,-2.25); 
  \end{tikzpicture}
  \vspace{-0.5cm}
  \caption{Image reconstruction of RU Lup for three different score-based priors. The first row on top shows from left to right the gridded visibilities (computed with our own gridding algorithm), the resulting dirty image obtained by applying an inverse Fourier transform on the gridded visibilities, the dirty image obtained with the \texttt{tclean} task (robust parameter of 0.0), CLEAN's model image and CLEAN's restored image. The rest of the figure are posterior samples and pixel-wise statistics (median and percentile range) from our approach for score-based priors trained under different SDE (either VE or VP) and different datasets (either SKIRT or PROBES). The residuals are computed using the leftmost posterior sample in each row.}
  \label{fig:rulup_3sbms}
\end{figure*}

\subsection{Prior Variability Testing}
We finally investigate how varying the prior affects the reconstructed images by focusing on RU Lup using three different score-based priors: two trained on the PROBES dataset using the VE and VP SDEs, and one trained on the SKIRT dataset with the VP SDE. In Figure \ref{fig:rulup_3sbms}, we present posterior samples, pixel-wise statistics, and residuals (computed for one representative posterior sample) with their corresponding $\chi^2$ values for Ru Lup under these different priors. While the overall posterior statistics vary, the main disk features remain consistent across priors, which points to their robustness to prior choice (given a sufficiently expressive prior). When the likelihood is highly informative, a sufficiently broad prior should not significantly influence regions of the posterior distribution where the data is very constraining. On the other hand, unconstrained regions, like unresolved components of the systems, should be prior-driven regions. These regions typically exhibit higher variability across posteriors samples, which is reflected in the percentile range (see Figure \ref{fig:all_disks} as well).

Additionally, we observe that the two VP SDE priors, despite being trained on different datasets, produce similar smooth structures in the disk's inner region. This suggests that in cases of prior misspecification, the VP SDE enforces a stronger prior, leading to slightly higher $\chi^2$ values. Conversely, posterior samples and statistics derived from the VE SDE exhibit richer substructures, implying that SBMs trained under this SDE may be better suited for inference tasks involving misspecified priors.  

\section{Conclusion}

In this work, we introduced a Bayesian approach to the synthesis imaging task in radio interferometry. By employing score-based models within the formalism of stochastic differential equations, we demonstrated that our approach constitutes a powerful framework for tackling the challenges of Bayesian inference in high-dimensional spaces. With a neural network serving as a highly expressive prior, our method efficiently samples the posterior, enabling the retrieval of realistic Bayesian uncertainties. Moreover, our approach is agnostic to the measured radio interferometric data and naturally incorporates the physics of the measurement process and the noise model. Even with a misaligned prior, our method achieves competitive performance when compared to state-of-the-art imaging algorithms, with the advantage of providing uncertainty quantification. 

Our framework has, however, some disadvantages compared to other imaging algorithms. First, it lacks scalability, as sampling from score-based models can be significantly slower for higher-dimensional data. Consequently, our current approach does not aim to replace state-of-the-art imaging algorithms for very large images, such as those required for wide-field imaging; instead, it presents a viable alternative for relatively small images. Secondly, while the variance preserving SDE sampling scheme can be calibrated using additional Predictor-Corrector steps as demonstrated through empirical coverage tests on simulations, we did not succeed in systematically achieving posterior calibration for variance exploding score models. We plan on conducting further testing of our inference process while relaxing some of the built-in assumptions in our current approach. This may reveal potential strategies for achieving unbiased uncertainty quantification with score-based priors for imaging algorithms.

\section*{Acknowledgements}
This research was made possible by a generous donation by Eric and Wendy Schmidt on the recommendation of the Schmidt Futures Foundation. We are grateful for useful discussions with Antoine Bourdin. The work is in part supported by computational resources provided by Calcul Quebec and the Digital Research Alliance of Canada. Y.H. and L.P. acknowledge support from the National Sciences and Engineering Research Council (NSERC) of Canada grant RGPIN-2020-05073 and 05102, the Fonds de recherche du Québec grant 2022-NC-301305 and 300397, and the Canada Research Chairs Program. The work of A.A. was partially funded by an NSERC CGS D scholarship. 
N.D. was supported through an Undergraduate Student Research Award provided by NSERC. 
M.B. gratefully acknowledges the support provided by Schmidt Sciences. M.B. \& A.M.M.S. gratefully acknowledge support from the UK Alan Turing Institute under grant reference EP/V030302/1.

Software used: \texttt{astropy} \citep{astropy:2013,astropy:2018}, \texttt{jupyter} \citep{jupyter}, \texttt{matplotlib} \citep{matplotlib} , \texttt{numpy} \citep{numpy}, \texttt{PyTorch} 
 \citep{pytorch}, \texttt{tqdm} \citep{tqdm}, \texttt{CASA} \citep{CASA}, \texttt{mpol} \citep{mpol}, \texttt{visread} \citep{visread}

\newpage
\bibliography{references}

\newpage
\appendix

\section{Derivation of the convolved likelihood approximation}\label{app:cla}

\subsection{Background}\label{sec:background}
In this appendix, we perform the derivation for the core approximation used in this work to make posterior sampling tractable, namely the convolved likelihood $p_t(\mathbf{y} \mid \mathbf{x})$ in equation \eqref{eq:convolved_likelihood}. 

In the context of continuous-time diffusion models introduced by \citep{Song_sbm_2021}, we work with the stochastic process $\mathbf{X}_t(\omega): \Omega \times [0, 1] \rightarrow \mathbb{R}^n$, defined on the measurable space $(\mathbb{R}^n, \mathcal{B}(\mathbb{R}^n), \{\mathcal{F}_t\}, \mathbf{W})$ where $\mathcal{B}(\mathbb{R}^n)$ is the Borel $\sigma$-algebra associated with $\mathbb{R}^n$, $\{\mathcal{F}_{t}\}$ is a set of filtrations and $\mathbf{W}$ is the Wiener measure. We will describe the stochastic process associated with the SDE of the random variable of interest $\mathbf{X}_t$ using the SDE perturbation kernel, $p_t(\mathbf{x}_t \mid \mathbf{x}_0)$, where $t \in [0, 1]$ is the time index of the SDE. More specifically, we consider the Gaussian perturbation kernel that correspond to the Variance-Preserving (VP) SDE \citep{Song_sbm_2021,Ho2020}
\begin{equation}
    p_t(\mathbf{x}_t \mid \mathbf{x}_0) = \mathcal{N}(\mathbf{x}_t \mid \mu(t) \mathbf{x}_0, \sigma^2(t) \bbone_{n \times n})\, .
\end{equation}
In this work, $\mathbf{x}_0$ is an image of the protoplanetary disk we wish to infer from the interferometric data. Since the kernel is Gaussian, we can express $\mathbf{x}_t$, a noisy image of the protoplanetary disk, directly in term of $\mathbf{x}_0$ and pure noise,
\begin{equation}\label{eq:reparametrization}
    \mathbf{x}_t = \mu(t)\mathbf{x}_0 + \sigma(t) \mathbf{z}\, ,
\end{equation}
where  $\mathbf{z} \sim \mathcal{N}(0, \bbone_{n \times n})$.

Since our goal is to sample from the posterior, we specify the $t=0$ boundary condition of the SDE as the posterior distribution $p(\mathbf{x}_0 \mid \mathbf{y})$. We can construct the marginal of the posterior at any time $t$ by applying Bayes' theorem, taking the logarithm and taking the gradient with respect to $\mathbf{x}_t$. However, the quantity $\grad_{\mathbf{x}_t} \log p(\mathbf{y} \mid \mathbf{x}_t)$, the second term on the RHS of equation \eqref{eq:score_bayes}, is intractable to compute since it involves an expectation over $p(\mathbf{x}_0 \mid \mathbf{x}_t)$
\begin{equation}
    p_t(\mathbf{y} \mid \mathbf{x}_t) = \int d\mathbf{x}_0\, p(\mathbf{y} \mid \mathbf{x}_0) p(\mathbf{x}_0 \mid \mathbf{x}_t)\, .
\end{equation}

To simplify this expression, we reverse the conditional $p(\mathbf{x}_0 \mid \mathbf{x}_t)$ using Bayes' theorem to get the known perturbation kernel of the SDE
\begin{equation}
    p_t(\mathbf{y} \mid \mathbf{x}_t) = \int d\mathbf{x}_0\, p(\mathbf{y} \mid \mathbf{x}_0) p(\mathbf{x}_t \mid \mathbf{x}_0) \frac{p(\mathbf{x}_0)}{p(\mathbf{x}_t)}\, .
\end{equation}

As mentionned in Sec. \ref{sec:cla}, the convolved likelihood approximation involves in treating the ratio $p(\mathbf{x}_0) / p(\mathbf{x}_t) $ as a constant or setting it to 1.

\subsection{Gaussian likelihood convolution}

We now evaluate the convolution between the likelihood and the perturbation kernel
\begin{equation}\label{eq:conv_lh}
    p_t(\mathbf{y} \mid \mathbf{x}_t) \approx \int d\mathbf{x}_0\, p(\mathbf{y} \mid \mathbf{x}_0) p(\mathbf{x}_t \mid \mathbf{x}_0)\, .
\end{equation}
The likelihood is a Gaussian distribution with covariance matrix $\Sigma \in \mathbb{R}^{2 m \times 2 m}$, measured empirically as described in section \ref{sec:data_and_phys},
\begin{equation}
    p(\mathbf{y} \mid \mathbf{x}_0) = \mathcal{N}(\mathbf{y} \mid A \mathbf{x}_0, \Sigma)\, .
\end{equation}
In what follows, we assume that the physical model, $A$, is not a singular matrix.  We define ${\tilde{A} \equiv S \mathcal{F} P_{\mathrm{beam}}} \in \mathbb{C}^{m \times n}$ to be the physical model. For the construction in this appendix to work, we redefine this complex matrix into its real-valued equivalent matrix
\begin{equation}
    A \equiv \begin{pmatrix}
        \Re(\tilde{A})  \\
        \Im(\tilde{A}) 
    \end{pmatrix}
    \in \mathbb{R}^{2m \times n}
\end{equation}
In this construction, every complex random variable, e.g., $\tilde{\bm{\eta}} \in \mathbb{C}^m$, is reformulated as an equivalent real-valued random variable
\begin{equation}
    \bm{\eta} = (\Re(\tilde{\bm{\eta}}), \Im(\tilde{\bm{\eta}})) \in \mathbb{R}^{2m}
\end{equation}

To evaluate the convolution in equation \eqref{eq:conv_lh}, we change the variables of integration for both the likelihood and the perturbation kernel. We first recall the data generating process, equation \eqref{eq:vis}, 
$
    \mathbf{y} = A \mathbf{x}_0 + \boldsymbol{\eta}\, ,
$
where $\mathbf{y} \equiv \mathcal{V}$ are the observed or simulated visibilities. We multiply equation \eqref{eq:vis} by $\mu(t)$, then use the reparameterization in equation \eqref{eq:reparametrization} to get
\begin{equation}\label{eq:rv_conv_lh}
    \boldsymbol{\eta}_t \equiv \mu(t) \mathbf{y} - A \mathbf{x}_t = \mu(t) \boldsymbol{\eta} - A \sigma(t) \mathbf{z}\, .
\end{equation} 
We now extract the form of the perturbation kernel from equation \eqref{eq:rv_conv_lh}. Recall that $\mathbf{z} \sim \mathcal{N}(0, \bbone_{n \times n})$ is a normally distributed random variable (see appendix \ref{sec:background}).  Using this, we can now obtain the perturbation kernel of the random variable $\boldsymbol{\eta}$
\begin{equation}
    p(\boldsymbol{\eta}_t \mid \boldsymbol{\eta}) =  \mathcal{N}(\boldsymbol{\eta}_t \mid \mu(t) \boldsymbol{\eta}, \sigma^2(t)\Gamma)\, ,
\end{equation}
where
\begin{equation}\label{eq:covariance_sigma}
    \Gamma \equiv A A^T 
    = \begin{pmatrix}
        \Re(\tilde{A})\Re(\tilde{A})^T & \Re(\tilde{A})\Im(\tilde{A})^T \\
        \Im(\tilde{A})\Re(\tilde{A})^T & \Im(\tilde{A})\Im(\tilde{A})^T
    \end{pmatrix}\, .
\end{equation}

We then rewrite equation \eqref{eq:conv_lh} in terms of $\boldsymbol{\eta}$ and $\boldsymbol{\eta}_t$. Using equation \eqref{eq:rv_conv_lh}, we can relate the kernel $p(\boldsymbol{\eta}_t \mid \boldsymbol{\eta})$ to $p(\mathbf{x}_t \mid \mathbf{x}_0)$ by making use of the change of variable formula for probability densities, we obtain
\begin{equation}
    p(\boldsymbol{\eta}_t \mid \boldsymbol{\eta}) 
        = \frac{1}{\lvert \det A \rvert }p(\mathbf{x}_t \mid \mathbf{x}_0)\, .
\end{equation}
This change of variable can only occur if $A$ is not singular.
By changing the variable of integration in equation \eqref{eq:conv_lh} to $\boldsymbol{\eta}$, we introduce another Jacobian determinant, $d \boldsymbol{\eta} = \lvert\det A\rvert d \mathbf{x}_0$. Finally, we make use of the equality $p(\mathbf{y} \mid \mathbf{x}_0) = p(\boldsymbol{\eta})$ to write
\begin{equation}
    p_t(\mathbf{y} \mid \mathbf{x}_t) \approx \int d\boldsymbol{\eta}\, p(\boldsymbol{\eta}) p(\boldsymbol{\eta}_t \mid \boldsymbol{\eta})\, . 
\end{equation}

Moreover, we note that the factor $\mu(t)$ scales the mean and covariance of the likelihood
\begin{equation}
    p(\mu(t) \boldsymbol{\eta}) 
    = \mathcal{N}(0, \mu^2(t) \Gamma)\, .
\end{equation}
We obtain the distribution of $\boldsymbol{\eta}_t$ by evaluating analytically the convolution implied by the sum of random variables on the RHS of equation \eqref{eq:rv_conv_lh}
\begin{equation}\label{eq:conv_lh_final}
    p(\boldsymbol{\eta}_t) = \mathcal{N}(\boldsymbol{\eta}_t  \mid 0, \mu^2(t) \Sigma + \sigma^2(t)\Gamma)\, ,
\end{equation}
which we can rewrite as 
\begin{equation}
    p(\mathbf{y} \mid \mathbf{x}_t) \approx 
        \mathcal{N}(\mu(t) \mathbf{y} \mid A\mathbf{x}_t, \mu^2(t) \Sigma + \sigma^2(t)\Gamma)\, .
\end{equation}
In the appendix that follows, we discuss in more detail the structure of the covariance matrix $\Gamma$.

\subsection{The convolved likelihood for imaging in Fourier space}\label{app:gamma_approx}

We now discuss in more detail the structure of the covariance $\Sigma$ and show how we arrive to the expression of equation \eqref{eq:convolved_likelihood} with $\Sigma_{jk} = \frac{1}{2}\delta_{jk} + \frac{1}{2} \delta_{j0}\delta_{k0}$. 
By choosing to model sky brightness $\mathbf{x}$ multiplied by the primary beam $P$, the structure of the covariance is mostly determined by the Fourier operator, $\mathcal{F}$. The effect of the sampling function, $S$, is only to select or remove rows from the covariance matrix. As such, we invoke this function only at the end of this section.
To make things simpler, we only consider the 1D Fourier operator in order to give an intuition of the behavior for the 2D case. This intuition will extend naturally to the 2D case. 

With these simplifications in place, we replace the physical model by the unitary Fourier operator, with elements
\begin{equation}
    \tilde{A}_{k \ell} = \frac{1}{\sqrt{n}}\omega^{k \ell} ,\qquad k,\ell \in \{0, \dots, n-1\}\,  .
\end{equation}
$\omega = e^{-2 \pi i / n}$ is the n\textsuperscript{th}-root of unity and $i \equiv \sqrt{-1}$ is the imaginary unit. We start by evaluating the elements of the first bloc of the covariance $\Gamma$ (see equation \eqref{eq:covariance_sigma}). Using Euler's formula and taking the real part of $\omega$, we get
\begin{equation}
    [\Re(\tilde{A})\Re(\tilde{A})^T]_{jk} = \frac{1}{n}
    \sum_{\ell = 0}^{n-1}  \cos\left(\frac{2 \pi j \ell}{n}\right)\cos\left(\frac{2 \pi k\ell}{n}\right)\, .
\end{equation}
By using the trigonometric identity $\cos(\alpha)\cos(\beta) = \frac{1}{2}(\cos(\alpha-\beta) + \cos(\alpha+\beta))$ and the general result for the sum of a geometric series, we obtain
\begin{align}
    \nonumber
    [\Re(\tilde{A})\Re(\tilde{A})^T]_{jk}
    &= \frac{1}{2n}\sum_{\ell = 0}^{n-1} \cos\left(\frac{2 \pi \ell (j - k)}{n}\right) + \frac{1}{2n}\sum_{\ell = 0}^{n-1} \cos\left(\frac{2 \pi \ell (j + k) }{n}\right) \\
    &= 
    \begin{cases}
        1, & j = k = 0 \mod n/2\\[0.5ex]
        \frac{1}{2}, & j = k \not= 0 \mod n/2 \\[0.5ex]
        \frac{1}{2}, & j + k = n \\
        0, & \text{otherwise}
    \end{cases}\, . \label{eq:block0}
\end{align}

We get a formula for the other blocs of the covariance matrix using similar arguments
\begin{align}
    [\Im(\tilde{A}) \Im(\tilde{A})^T]_{jk} &= 
    \begin{cases}
        \frac{1}{2}, & j = k \not= 0 \mod n/2 \label{eq:block_1}\\
        -\frac{1}{2}, & j + k = n\\
        0, & \text{otherwise}
    \end{cases}\, ,\\
    [\Re(\tilde{A}) \Im(\tilde{A})^T]_{jk} &= 0 \, ,\\
    [\Im(\tilde{A}) \Re(\tilde{A})^T]_{jk} &= 0 \label{eq:block_4} \, .
\end{align}

\begin{figure*}[!ht]
    \centering
    \vspace{-0.5cm}
    \begin{tikzpicture}
        \centering
        \node (image1) at (0,0){\includegraphics[scale = 0.45]{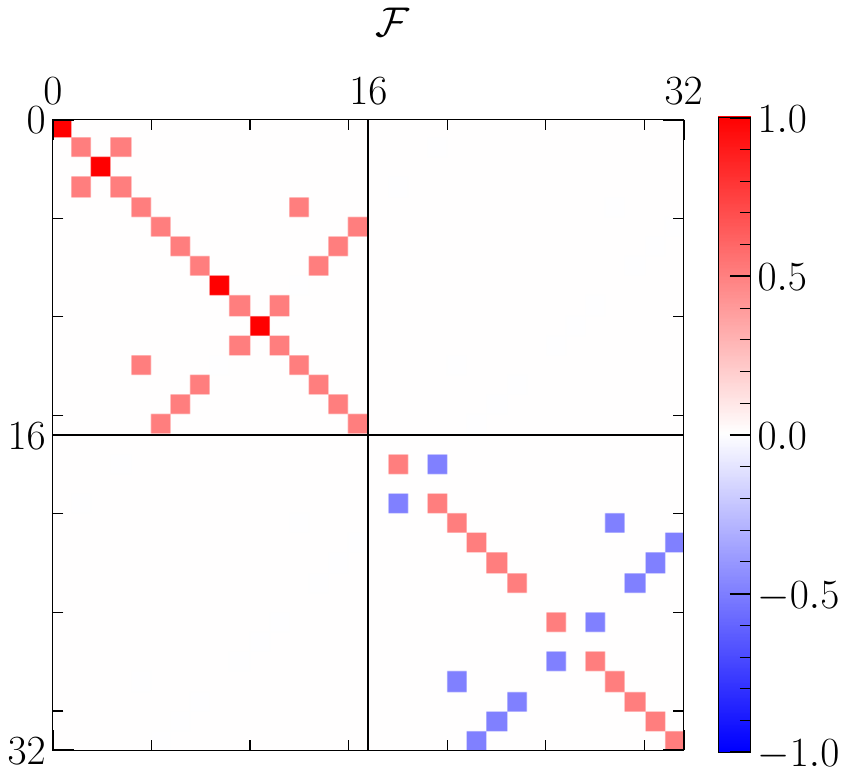}}; 
        \node (image2) at (8,0){\includegraphics[scale = 0.45]{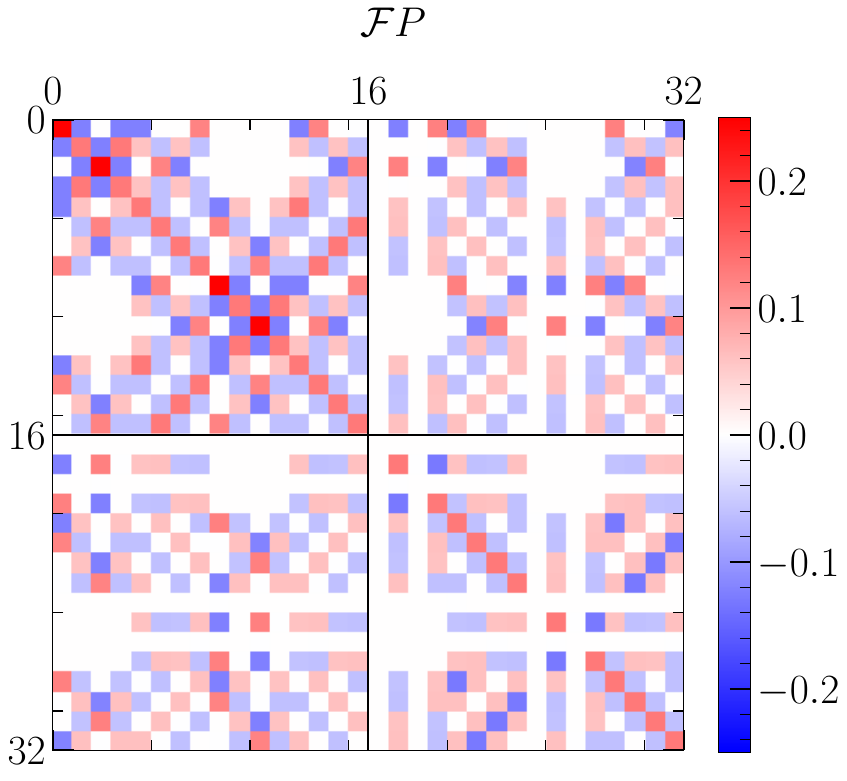}};
    \end{tikzpicture}
    \caption{The covariance matrix $\Gamma$ without including the primary beam in our forward model $\tilde{A}$ (left) and including it in the forward model (right) for an input in $\mathbb{R}^{(4\times 4)}$. On the left, the structure of the covariance is greatly simplified since it is entirely determined by the 2D Fourier operator $\mathcal{F}$.}
    \label{fig:cov_gamma}
\end{figure*}

In summary, we obtain a block-diagonal structure for the covariance. The diagonal components of the real and imaginary blocks are equal to $1/2$, except for their $j=k=0$ components (sometimes called the DC component) and their component at the Nyquist frequency, $j = k = n/2$. In Figure \ref{fig:cov_gamma}, we present the covariance matrix associated with the 2D Fourier operator for $n=4\times4$ and illustrate how the inclusion of the primary beam in the forward model increases the complexity of the covariance. As before, the diagonal components equal $1/2$ with special cases for the DC and Nyquist components (3 per block in total). Since our images are real-valued, the covariance is unity for the real DC component and null for the imaginary DC component. The null entries of the covariance can be removed with the sampling function since real vectors do not contribute to these Fourier components. As it turns out, the sampling function considered in this work also removes the Nyquist frequencies. This is to say that the size of the pixels considered in section \ref{sec:results} over-samples the signal in the data. As such, these cases can safely be ignored in our final expression. 

To speed up computation, we simplify further the covariance matrix by ignoring off-diagonal terms. This practical solution has little effects on the results in this work. Thus, our final expression for the covariance is $\Gamma_{jk} = \frac{1}{2}\delta_{jk} + \frac{1}{2}\delta_{j0}\delta_{k0}$, where $\delta_{jk}$ denotes the Kronecker delta.

\section{Additional posterior samples}\label{app:method_comparison}
We show additional posterior samples from our approach for the disks HD 143006 and WaOph 6 whose pixel-wise statistics computed with our apporoach were presented in Figure \ref{fig:method_comparison}.
\begin{figure}[!ht]
    \centering
    \includegraphics[width=0.9\textwidth]{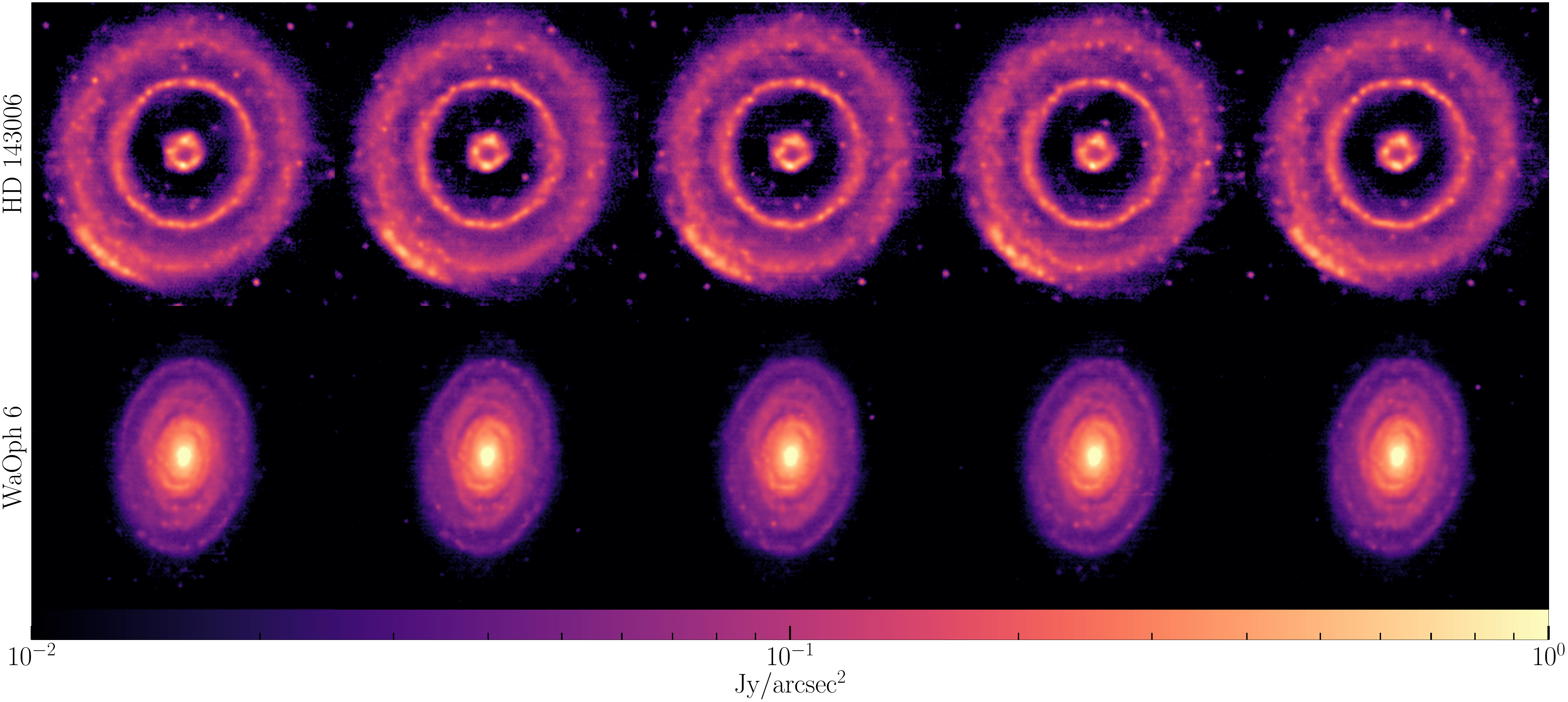}
    \vspace{-0.2cm}
    \caption{Posterior samples of the disks HD143006 (first row) and WaOph 6 (second row) presented in Figure \ref{fig:method_comparison}.}
    \label{fig:additional_samples}
\end{figure}

\begin{figure*}[!ht]
    \centering
    \includegraphics[scale=0.18]{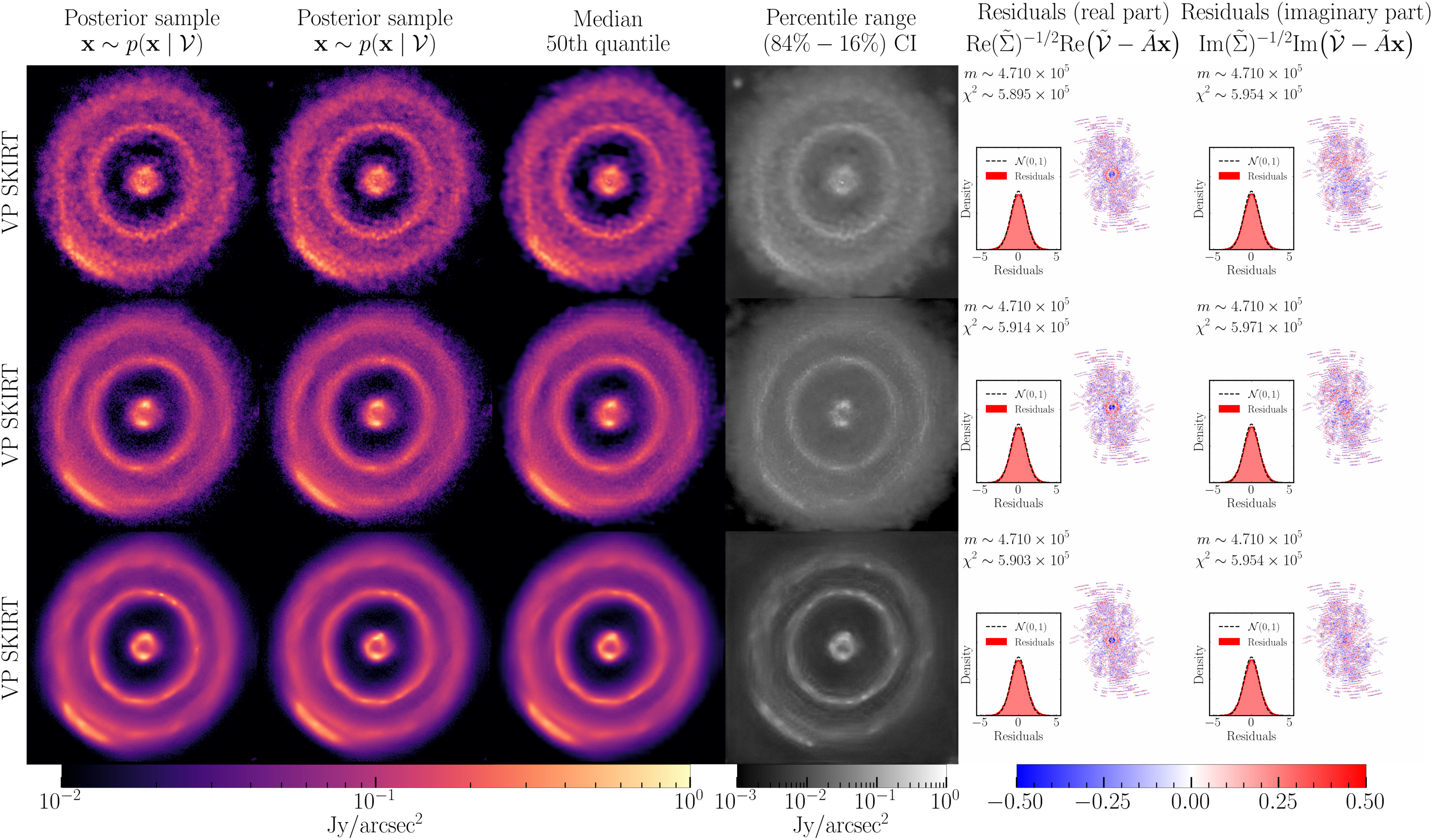}
    \caption{Effect of the scaling parameter $s$ on posterior samples, pixel-wise statistics (median and percentile range) and residuals for the disk HD 143006. Each row corresponds to a different scaling parameter.}
    \label{fig:scaling_factor_comparison}
\end{figure*}

\section{Scaling factor effect}\label{app:scale_factor}
We show Figure \ref{fig:scaling_factor_comparison} the effect of the scaling parameter $s$ on the posterior samples of the disk HD 143006 using the VP SKIRT prior. We choose this prior for this test for its ability to capture a wider dynamic range compared to the PROBES. This allows us to distinguish how some structures are encoded at higher/smaller intensities in this prior.

\section{Sampler comparison on DSHARP data}\label{app:sampler_comparison}
\begin{figure*}[!ht]
    \centering
    \includegraphics[scale=0.18]{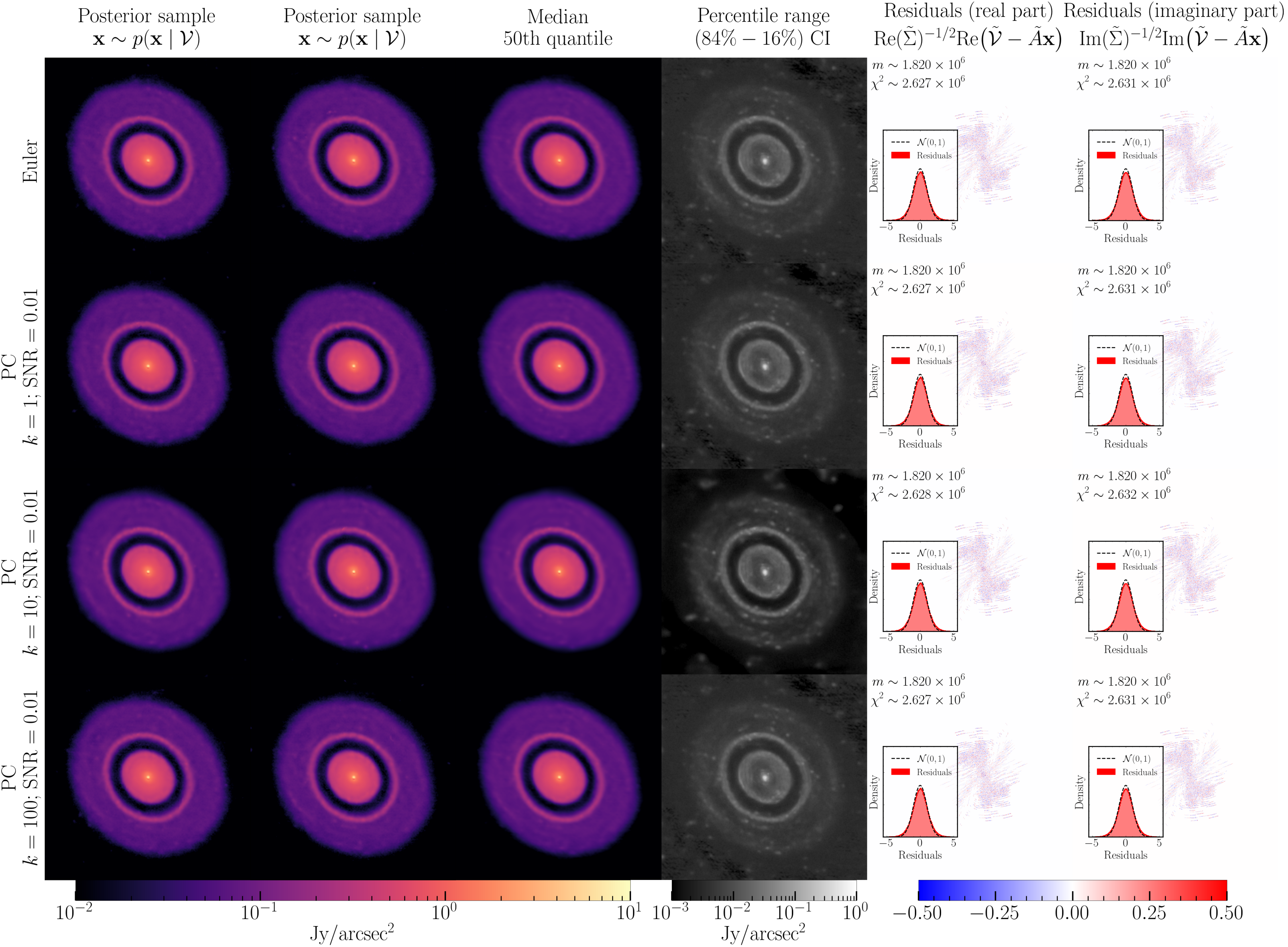}
    \caption{Comparison of Euler sampler (first row) with different combinations of corrector and SNR parameters for the PC sampler using the VE PROBES prior. $\chi^2$ values are approximately the same across samplers.}
    \label{fig:sampler_comparison}
\end{figure*}

We compare Figure \ref{fig:sampler_comparison} the Euler-Maruyama sampler to different combinations of PC parameters (in a similar way as we did on simulations) for the disk Elias 24. We operate with $4000$ predictor steps in all cases and choose smaller SNR values than on simulations as we experimented some instabilities during the sampling procedure for higher SNR. Since it is not possible to run TARP tests on real data, we only show here $\chi^2$ results. We observe that $\chi^2$ values are approximately the same when using Predictor-Corrector sampler compared to Euler sampler on real data using the galaxy PROBES prior. 

\newpage
\section{Stochastic differential equations for score-based models}\label{app:sdes}
For completeness, we specify here the different functions defining the VE and the VP SDE Table \ref{tab:sde_params}. We strongly recommend the work of \cite{Sarkka2019} and \cite{Song_sbm_2021} for more details on the formalism of score-based models and SDEs.
\begin{table*}[!ht]
\centering
\begin{tabular}{ccccc}
\hline
SDE & $\beta(t)$ & $g(t)$ & $\mu(t)$ & $\sigma(t)$ \\
\hline
VP  & $\beta_{\min} + t(\beta_{\max} - \beta_{\min})$ & $\sqrt{\beta(t)}$  & $\exp\left({-\frac{1}{2}\int_{0}^{t} \beta(s)ds}\right)$ & \big[$1-\exp\big(-\int_{0}^{t} \beta(s)ds\big)\big]^{1/2}$ \\ 
VE  & 0 & $\displaystyle \sqrt{\frac{d[\sigma^2 (t)]}{dt}}$ & $1$ & $\left(\sigma_{\text{max}}/\sigma_{
\text{min}}\right)^{t} \sigma_{\text{min}}$\vspace{0.2cm}\\ 
\hline
\end{tabular}
\label{tab:sde_params}
\caption{From left to right, the drift coefficient $\beta(t)$, the diffusion coefficient $g(t)$, and the mean and standard deviation of the Gaussian perturbation kernel $p_{t}(\mathbf{x}_t\mid \mathbf{x}_0)$ for the VP (first row) and the VE (second row) SDEs. The choice for the pairs of parameters $(\sigma_{\min}, \sigma_{\max})$ for the VE SDE and $(\beta_{\min},\beta_{\max})$ for the VP SDE depends on the dataset used for training.}
\end{table*}
\end{document}